%% file: paper.tex
\pdfoutput=1

\documentclass[twocolappendix,apj]{emulateapj}
\usepackage{apjfonts}
\usepackage[colorlinks=true,urlcolor=blue,linkcolor=blue,citecolor=blue]{hyperref}
\usepackage{amsmath}
\usepackage{natbib}
\usepackage{graphicx}

\input{citation_fix}
\graphicspath{{./}}

\def\Mnsf{{M}_{\rm nsf}}
\def\Msf{{M}_{\rm sf}}
\def\Mg{{M}_{\rm g}}
\def\Mst{{M}_\star}
\def\MH2{M_{\rm H_2}}

\def\SFR{\dot{M}_\star}

\def\Fsf{F_{\rm sf}}
\def\Fp{F_{\rm +}}
\def\Fm{F_{\rm -}}
\def\Fmdyn{F_{\rm -,d}}
\def\Fmfb{F_{\rm -,fb}}

\def\rhoSFR{{\dot{\rho}}_\star}
\def\rhoH2{\rho_{\rm H_2}}
\def\epsff{\epsilon_{\rm ff}}
\def\epsint{\epsilon_{\rm int}}
\def\avir{\alpha_{\rm vir}}
\def\avirsf{\alpha_{\rm vir,sf}}
\def\cs{c_{\rm s}}
\def\st{\sigma_{\rm t}}
\def\stot{\sigma_{\rm tot}}

\def\Sg{\Sigma_{\rm g}}
\def\SH2{\Sigma_{\rm H_2}}
\def\Ssf{\Sigma_{\rm sf}}
\def\SSFR{\dot\Sigma_\star}

\def\fsf{f_{\rm sf}}
\def\fH2{f_{\rm H_2}}
\def\fgas{f_{\rm g}}

\def\Nc{N_{\rm c}}
\def\tsf{t_{\rm sf}}
\def\tnsf{t_{\rm nsf}}
\def\tglob{\tau_{\rm dep}}
\def\tH2{\tau_{\rm dep,H_2}}
\def\tdep{t_\star}
\def\taust{\tau_\star}
\def\tp{\tau_{+}}
\def\tm{\tau_{-}}
\def\tmd{\tau_{\rm -,d}}
\def\tmfb{\tau_{\rm -,fb}}
\def\tff{t_{\rm ff}}

\def\tsfr{\Delta t}
\def\tsfe{\tau_{\rm e,sf}}

\def\pc{{\rm~pc}}
\def\kpc{{\rm~kpc}}
\def\Myr{{\rm~Myr}}
\def\Gyr{{\rm~Gyr}}
\def\Msun{{\rm~M_\odot}}
\def\Msunyr{{\rm~M_\odot~yr^{-1}}}
\def\Msunpc2{{\rm~M_\odot~pc^{-2}}}
\def\kms{{\rm~km~s^{-1}}}
\def\cc{{\rm~cm^{-3}}}
\def\K{{\rm~K}}

\begin{document}


\shorttitle{The origin of long gas depletion times}
\shortauthors{Semenov, Kravtsov \& Gnedin}
\slugcomment{Accepted for publication in The Astrophysical Journal} 

\title{The physical origin of long gas depletion times in galaxies}

\author{Vadim A. Semenov\altaffilmark{1,2,$\star$}, Andrey V. Kravtsov\altaffilmark{1,2,3} and Nickolay Y. Gnedin\altaffilmark{1,2,4}}


\altaffiltext{1}{Department of Astronomy \& Astrophysics, The University of Chicago, Chicago, IL 60637 USA}
\altaffiltext{2}{Kavli Institute for Cosmological Physics, The University of Chicago, Chicago, IL 60637 USA}
\altaffiltext{3}{Enrico Fermi Institute, The University of Chicago, Chicago, IL 60637 USA}
\altaffiltext{4}{Fermilab Center for Particle Astrophysics, Fermi National Accelerator Laboratory, Batavia, IL 60510-0500 USA}
\altaffiltext{$\star$}{semenov@uchicago.edu}


\begin{abstract}
We present a model that explains why galaxies form stars on a time scale significantly longer than the time scales of processes governing the evolution of interstellar gas. We show that gas evolves from a non-star-forming to a star-forming state on a relatively short time scale and thus the rate of this evolution does not limit the star formation rate. Instead, the star formation rate is limited because only a small fraction of star-forming gas is converted into stars before star-forming regions are dispersed by feedback and dynamical processes. Thus, gas cycles into and out of star-forming state multiple times, which results in a long time scale on which galaxies convert gas into stars. Our model does not rely on the assumption of equilibrium and can be used to interpret trends of depletion times with the properties of observed galaxies and the parameters of star formation and feedback recipes in simulations. In particular, the model explains how feedback self-regulates the star formation rate in simulations and makes it insensitive to the local star formation efficiency. We illustrate our model using the results of an isolated $L_*$-sized galaxy simulation that reproduces the observed Kennicutt-Schmidt relation for both molecular and atomic gas. Interestingly, the relation for molecular gas is almost linear on kiloparsec scales, although a nonlinear relation is adopted in simulation cells. We discuss how a linear relation emerges from non-self-similar scaling of the gas density PDF with the average gas surface density.
\end{abstract}

\section{Introduction}
\label{sec:introduction}
\setcounter{footnote}{0}

One of the widely recognized basic facts about observed galaxies is that they convert gas into stars inefficiently. Star formation rates of galaxies are surprisingly low, given the total mass and density of their interstellar gas and the expected time scales of processes driving star formation. 

As an example, the star formation rate (SFR) of the Milky Way (MW) is $\SFR \sim 1-2 \Msunyr$, while its gas mass is $\Mg \sim 10^{10}\ \Msun$. Thus, the time scale at which the Galaxy would deplete its gas while forming stars at the current rate is $\tglob \equiv \Mg/\SFR \sim 5-10\Gyr$. The depletion time scales of a population of normal star-forming galaxies are comparable and span a range of $\sim 2-10$ Gyr \citep{Kennicutt.1989,Kennicutt.1998,Bigiel.etal.2008}. The denser molecular phase of the interstellar medium (ISM) is depleted on a similarly long time scale of $\tH2 \equiv \MH2/\SFR \sim 1-3$ Gyrs \citep{Kennicutt.1989,Kennicutt.1998,wong_blitz02,Bigiel.etal.2008,leroy_etal08,Leroy.etal.2013}. 

Compared to the time scales of any dynamical processes that are potentially relevant for star formation, the observed gas depletion times are very long indeed. For example, the orbital period of gas at the solar radius is $t_{\rm orb} \sim 200$ Myr, and the MW is thus depleting its gas on the timescale of $\sim 25-50$ such periods. On average, galaxies deplete their gas on a timescale of $\sim 10-20$ orbital periods \citep[][]{Kennicutt.1998,wong_blitz02,leroy_etal08,daddi_etal10}.

The orbital period, $t_{\rm orb}$, is probably the longest of the relevant dynamical timescales one can think of. For example, the turbulent crossing time is usually $t_{\rm cross}=h/\sigma \sim 10-30\Myr$, where $\sigma\gtrsim 10 \kms$ is the velocity dispersion of gas in galactic disks and $h\sim 100-300\pc$ is the disk scale height in the inner regions of galaxies. The free-fall time at the mean or midplane density, $\rho_0$, of galaxies spans a similar range: $t_{\rm ff,0}\equiv\sqrt{3\pi/32G\rho_0}\sim 10-50\Myr$. The timescale of molecular cloud collisions is $\lesssim 20\Myr$ \citep[e.g.,][]{Tan.2000}. A given gas mass encounters a spiral arm on a timescale of $t_{\rm arm}\sim 2 \pi /(m[\Omega(R)-\Omega_{\rm p}])$, where $\Omega(R)=V_{\rm rot}/R$ is the angular frequency of gas rotation, $\Omega_{\rm p}$ is the pattern speed of spiral arms, and $m$ is the number of spiral arms. This timescale is $t_{\rm arm}\sim 50-200\Myr$, if we assume $\Omega_{\rm p} \sim 20 \rm \kms\,kpc^{-1}$ \citep[e.g.,][]{Bissantz.etal.2003}, $m \sim 2-4$ \citep[e.g.,][]{Davis.etal.2015} and $V_{\rm rot} \sim 220 \kms$ typically derived for MW-like galaxies. Numerical simulations of gaseous galactic disks show that star-forming molecular clouds may form on even shorter timescales of a few tens of Myrs \citep{Dobbs.etal.2012,Dobbs.etal.2015}.

In addition to being slow on global galactic scales, star formation is inefficient even in dense molecular star-forming regions, which convert only $\lesssim 1-10\%$ of gas into stars per local free-fall time \citep{Zuckerman.Evans.1974,zuckerman_palmer74,Krumholz.Tan.2007,Krumholz.etal.2012,Evans.etal.2014,Lee.etal.2016,Heyer.et.al.2016}. Such low efficiency arises because only $\sim 0.1-10\%$ of the dense gas is self-gravitating and collapsing into stars \citep{Froebrich.Rowles.2010}. 

However, the inefficiency of star-forming regions alone cannot explain long global depletion times. The \textit{local} depletion time in observed star-forming regions is $\tdep \sim 40-500 \Myr$ \citep[e.g.,][]{Evans.etal.2009,Evans.etal.2014,Lada.etal.2010,Lada.etal.2012,Heiderman.etal.2010,Gutermuth.etal.2011,Schruba.etal.2017}. Thus, although the scatter is significant, typical values of $\tdep$ are considerably smaller than the \textit{global} depletion time of molecular gas, $\tH2 \sim 1-3 \Gyr$. 

The large scatter in depletion times measured on small scales and the difference between local and global depletion time values indicate that only a fraction of molecular gas is actively forming stars at any given moment. Indeed, the global depletion time can be expressed as
\begin{equation}
\label{eq:tglob_def}
\tglob \equiv \frac{\Mg}{\SFR} = \frac{\taust}{\fsf},
\end{equation} 
where $\Mg$ is the total gas mass of the galaxy; $\taust \equiv \Msf/\SFR = \langle 1/\tdep \rangle_{\rm sf}^{-1}$ is the mass-weighted average over the depletion time distribution in star-forming regions, $\tdep$; and $\fsf \equiv \Msf/\Mg$ is the gas mass fraction in actively star-forming regions. A similar expression can be written for the global depletion time of molecular gas, $\tH2$, via a corresponding star-forming fraction $f_{\rm sf,H_2} \equiv \Msf/\MH2$. 

Thus, the depletion time measured on larger scales is longer than that in star-forming regions because, as the scale increases, more of non-star-forming gas is incorporated in the gas mass estimate. Likewise, when depletion time is estimated on larger scales, the scatter in $\tglob$ decreases as we average over  the distribution of local $\tdep$. This is indeed observed \citep{schruba_etal10,Schruba.etal.2017}, although some of the obtained variation may be due to observational effects \citep{Feldmann.etal.2011,Kruijssen.Longmore.2014}. 
 
The small scatter in the global depletion time in observed galaxies is manifested in the Kennicutt-Schmidt relation (KSR) between the surface density of gas and the star formation rate \citep[][see also \citealt{Sanduleak.1969,Madore.etal.1974}]{Schmidt.1959,Kennicutt.1989,Kennicutt.1998}. The relation is particularly tight and close to linear when only molecular hydrogen is used to estimate the surface density of gas \citep{wong_blitz02,Bigiel.etal.2008,Leroy.etal.2013}. Thus, a comprehensive model for the global depletion time must explain not only its value, i.e., the normalization of the KSR, but also both the scatter and the shape of the KSR on different scales.  

A number of useful global star formation frameworks and models have been developed over the last three decades \citep[e.g.,][]{Wyse.Silk.1989,Silk.1997,Tan.2000,Elmegreen.2002,Krumholz.McKee.2005,Li.etal.2005,
Krumholz.Thompson.2007,Saitoh.etal.2008,Krumholz.etal.2009,Silk.Norman.2009,
Ostriker.et.al.2010,Ostriker.Shetty.2011,Renaud.etal.2012,FaucherGiguere.etal.2013,
Federrath.2013,Elmegreen.2015,Salim.etal.2015}. Many of these models consider the physical processes shaping the form of the KSR while treating its normalization as a flexible constant. Other models also consider the physical origin of the normalization and long depletion timescale. 

One class of the latter models associates long depletion times with the fraction of gas in dense, self-gravitating regions of cold, supersonic molecular clouds with the log-normal gas density PDF \citep{Elmegreen.2002,Krumholz.McKee.2005,Krumholz.etal.2012}. Such models, however, assume that all of the molecular gas is in ``virialized'' star-forming molecular clouds and that the star formation efficiency in these clouds sets the global depletion time. This assumption, which has also been frequently adopted in galaxy simulations \citep[e.g.,][]{Robertson.Kravtsov.2008,Gnedin.etal.2009,Christensen.etal.2012,Kuhlen.etal.2012}, is at odds with a growing number of observations indicating that the depletion time of \textit{star-forming} molecular gas is in general considerably shorter than the global depletion time of \textit{all} molecular gas, $\tH2$, estimated on $\gtrsim $~kpc scales. Moreover, models and simulations of star formation in supersonic turbulent clouds show that the local efficiency of star formation is primarily a strong function of the virial parameter of the region, not just its density, temperature, and molecular fraction \citep[e.g.,][]{Krumholz.McKee.2005,Padoan.etal.2012,Padoan.etal.2017}, while the virial parameter can span a wide range of values \citep{Dobbs.etal.2011,Semenov.etal.2016}. 

Some models derive the Kennicutt-Schmidt relation and its normalization by assuming that stellar feedback regulates ISM turbulence so as to maintain vertical and/or \citet{Toomre.1964} equilibrium within gaseous disks  \citep{Ostriker.Shetty.2011,FaucherGiguere.etal.2013,Hayward.Hopkins.2017}. However, it is not clear a priori why equilibrium can generically be expected in galaxies as a whole or in kiloparsec-scale patches and why the star formation rate does not instead reach values at which gas is driven out in a wind. Moreover, it is still debated whether the turbulence within galactic disks is mainly driven by stellar feedback or by gravitational instabilities \citep[e.g.,][]{Krumholz.Burkhart.2016}. 

\citet{Saitoh.etal.2008} argued that SFR is controlled by the rate at which gas is supplied from the general ISM to the star-forming state, which makes it insensitive to the local efficiency of star formation. However, these authors measured the timescale at which gas is supplied to the star-forming state to be $\sim 100$ Myr and did not explain how this timescale relates to the much longer observed depletion times of total gas, $\tglob \sim 2-10$ Gyrs. 

In this paper, we aim to clarify the origin of the observed long gas depletion timescale in galaxies, taking into account both the inefficiency of star formation in star-forming clouds and the fact that not all of the molecular gas is actively forming stars. To this end, in Section \ref{sec:model} we present a simple conceptual framework that views the ISM as a highly dynamic medium in which gas evolves between non-star-forming and star-forming states. The key aspect of our framework is that it considers global gas depletion as a result of gas evolution that is driven by processes with associated characteristic timescales. This approach is conceptually similar to the framework of \citet{Kruijssen.Longmore.2014}, developed to explain and interpret the scatter and possible biases in observational measurements of the Kennicutt-Schmidt relation on different scales.

We illustrate our framework using a realistic simulation of a galactic disk that reproduces the observed depletion time and the Kennicutt-Schmidt relation. We describe the simulation in Section~\ref{sec:simulations}. In Section~\ref{sec:results} we demonstrate that the long global depletion time originates from the rapid cycling of ISM gas between non-star-forming and star-forming states on timescales $\sim 20-100$ Myr, in accord with the above estimates and the results of previous galactic disk simulations. On each such cycle, only a small fraction of the gas mass is converted into stars and thus $\tglob$ is long because a large number of such cycles would be required to deplete all available gas. We analyze the processes driving the rapid gas evolution and also use simulations to shed light on the reason why the depletion time of molecular gas is nearly independent of the gas surface density. We discuss our results and summarize our conclusions in Sections~\ref{sec:discussion} and \ref{sec:conclusions}.

\section{A model for gas depletion time}
\label{sec:model}

\begin{figure}
\centering
\includegraphics[width=\columnwidth]{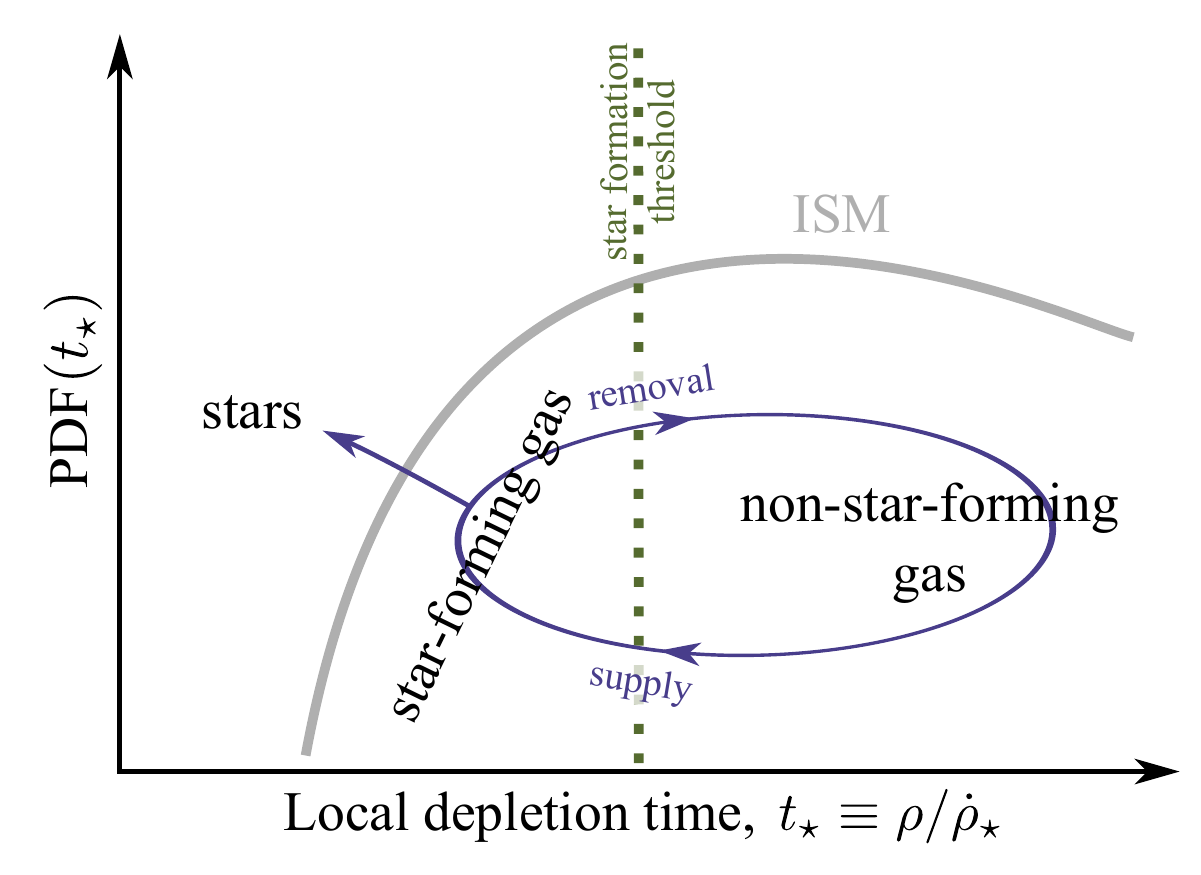}
\caption{\label{fig:pict} Schematic illustration of ISM gas evolution between non-star-forming and star-forming states. The thick gray line indicates the mass-weighted PDF of local gas depletion times, $\tdep \equiv \rho/\rhoSFR$, where $\rho$ and $\rhoSFR$ are the local densities of gas and SFR. The vertical dotted line corresponds to the threshold depletion time, $t_{\rm \star,max}$, separating star-forming and non-star-forming gas. The blue loop illustrates cycling of a gas parcel between these states under the influence of dynamical and feedback processes that supply and remove star-forming gas. }
\end{figure}

The interstellar gas in galaxies is a multiphase, dynamic medium spanning several orders of magnitude in density and temperature. To get a sense of the processes affecting the gas evolution in such a medium, we consider the evolution of individual gas parcels, massless tracers of gas flows in the ISM. One can think of a representative set of the ISM atoms as such tracers. At any given time, the local environment of such tracers can be estimated by averaging gas properties on some scale $l$ around the position of each tracer. The gas around tracers will expand and contract under the influence of dynamical processes, such as turbulence driven by gravitational instabilities and stellar feedback. Therefore, during evolution over a sufficiently long timescale, the conditions around each gas parcel can evolve between the states of long and short depletion time, $t_\star$, once or over many transition cycles. Such cycling of a gas parcel is schematically shown in Figure~\ref{fig:pict}. 

The probability density per unit time for a parcel to be converted into a star is given by $1/t_\star$ and one can define the depletion time for a single parcel as the time required for the integrated probability to reach unity.  In what follows, for conceptual simplicity, we will adopt a sharp threshold, $t_{\star,\rm max}$, separating non-star-forming, $t_\star>t_{\star,\rm max}$, and actively star-forming, $t_\star<t_{\star,\rm max}$, gas states. For a given distribution of $t_\star$, this threshold can be chosen in such a way that regions with $t_\star<t_{\star,\rm max}$ include most of the total star formation. 

The duration of a single cycle between the successive stages when the parcel's environment is in the star-forming state equals to the sum of the time spent in the non-star-forming, $\tnsf$, and star-forming, $\tsf$, stages. If we denote the average depletion time of the parcel during the star-forming stage as $\taust$, the parcel will have a probability of $\tsf/\taust$ to be converted into a star during the entire cycle. In other words, $\Nc=\taust/\tsf$ such cycles would be needed for the parcel to be incorporated into a star. Its depletion time can thus be written as
\begin{equation}
\label{eq:tglob1} 
t_{\rm dep} = \Nc(\tnsf + \tsf) = \Nc\tnsf + \taust=\left(\frac{\tnsf}{\tsf} + 1\right)\,\taust.
\end{equation}
Hence, the depletion time of a gas parcel is always longer than $\taust$ and it can be long because star formation during $\tsf$ is inherently inefficient, i.e.,  $\taust$ is long, and/or because only a small fraction of the evolution cycle is spent in the star-forming state, i.e., $\tnsf/\tsf$ is large. 

A group of parcels in a patch of the ISM has a distribution of $\taust$, $\tsf$, $\tnsf$ and the collective depletion time of the group is the average $\tglob=\langle 1/t_{\rm dep}\rangle^{-1}$ over these distributions. It is clear that if the depletion times of individual parcels are long, $\tglob$ will also be long. 

In practice, the $\tglob$ of an ISM patch is estimated from the instantaneous mass of gas, $\Mg$, and young stars formed over a time interval $\tsfr$, $\Mst(<\tsfr)$, such that the average star formation rate is $\langle\SFR\rangle_{\tsfr} = \Mst(<\tsfr)/\tsfr$ and the depletion time is defined as $\tglob \equiv \Mg/\langle\SFR\rangle_{\tsfr}$. This estimate of $\tglob$ can be related to the dynamics of individual gas parcels within the patch by noting that due to mass conservation, the instantaneous total mass of the gas parcels in the star-forming state, $\Msf$, evolves as $\dot{M}_{\rm sf}=\Fsf - \SFR$, where $\Fsf$ is the net instantaneous flux of the gas parcels through the star formation threshold and $\SFR$ is the instantaneous SFR of all parcels. After averaging this expression over the time interval $\tsfr$, we get
\begin{equation}
\label{eq:tglob_Fsf}
\tglob \equiv \frac{\Mg}{\langle \SFR \rangle_{\tsfr}} = \frac{\Mg}{\langle\Fsf\rangle_{\tsfr} - \langle\dot{M}_{\rm sf}\rangle_{\tsfr}}.
\end{equation}  

For brevity, in the following derivation, we omit explicit averaging, $\langle...\rangle_{\tsfr}$, but assume all fluxes and rates to be averaged over $\tsfr$.

In general, the average net flux of gas through the star formation threshold can be decomposed into positive and negative contributions,
\begin{equation}
\label{eq:Fsf}
\Fsf = \Fp - \Fm,
\end{equation}
which correspond to the supply and removal of star-forming gas illustrated in Figure~\ref{fig:pict}.

The positive flux $\Fp$ is controlled by a combination of global dynamical processes, e.g., gravitational instabilities, turbulence, cooling, etc., with a significant contribution from stellar feedback. The latter comes in the form of turbulence stirred by interactions of supernova-driven bubbles and by fountain outflows of gas from star-forming regions. The negative flux $\Fm = \Fmfb + \Fmdyn$ results from the destruction of star-forming regions both by feedback from young stars formed inside the regions, $\Fmfb$, and by shearing due to large-scale turbulence or differential rotation, expansion of gas behind galactic spiral arms, and other dynamical processes not directly related to star formation inside the regions, $\Fmdyn$. All fluxes can be parameterized with the characteristic timescales, i.e., $\tp$, $\tmfb$ and $\tmd$, on which gas is supplied to and removed from the star-forming state by the corresponding processes,
\begin{align}
\label{eq:Fc}
\Fp &\equiv \frac{\Mnsf}{\tp} = \Mg \frac{1-\fsf}{\tp},\\
\label{eq:Fm}
\Fm &\equiv \frac{\Msf}{\tm}  = \Fmfb + \Fmdyn=\Mg\fsf\left( \frac{1}{\tmfb} + \frac{1}{\tmd}\right),
\end{align} 
where $\fsf \equiv \Msf/\Mg$ is the star-forming mass fraction.

To make the relation between star formation and stellar feedback explicit, we can also parameterize $\Fmfb$ in a way similar to the parameterization of the mass outflow rate of feedback-driven galactic winds,
\begin{equation}
\label{eq:Ffb}
\Fmfb \equiv \xi \SFR = \Mg\fsf \frac{\xi}{\taust},
\end{equation}
where $\xi$ is the {\it mass-loading factor} and we used the definition of the average depletion time of {\it star-forming gas}, $\taust \equiv \Msf/\SFR$. In the context of Equation~(\ref{eq:Fm}) the mass-loading factor can also be interpreted as a relative rate of gas removal by feedback compared to the rate of star formation, i.e., $\xi \equiv \taust/\tmfb$.

An imbalance between the net gas flux into the star-forming state, $\Fsf$, and the average SFR may result in the evolution of the star-forming mass, which we also parameterize with the characteristic timescale, $\tsfe$:
\begin{equation}
\label{eq:tsfe}
|\dot{M}_{\rm sf}| \equiv \frac{\Msf}{\tsfe}.
\end{equation}

The final expression for the global depletion time can be readily derived by substituting Equations~(\ref{eq:Fsf}-\ref{eq:tsfe}) and $\fsf = \taust/\tglob$ into Equation~(\ref{eq:tglob_Fsf}),
\begin{equation}
\label{eq:tglob_xi} 
\tglob = \left(1+\xi+\frac{\taust}{\tmd} \pm \frac{\taust}{\tsfe}\right)\tp + \taust,
\end{equation}
where the sign in front of $\taust/\tsfe$ reflects the sign of $\dot{M}_{\rm sf}$.

If we compare the terms in this equation with those in Equation~(\ref{eq:tglob1}) for the depletion time of a single gas parcel, $t_{\rm dep} = \Nc \tnsf + \taust$, their physical meaning becomes clear. The timescale $\tp$ is analogous to the time $\tnsf$ that a gas parcel spends in the non-star-forming state, while the expression in parentheses is analogous to $\Nc=\taust/\tsf$, i.e., the average number of evolution cycles it would take for a single parcel to deplete its gas. Indeed,  Equation (\ref{eq:Fc}) gives $\Mnsf=\Fp\tp$, which means that $\tp$ is the time over which all of the non-star-forming gas will reach the star-forming state. Thus, $\tp$ is analogous to the average $\tnsf$ timescale for a collection of parcels. Likewise, the average rate at which the gas mass in the star-forming state is decreasing due to star formation, dispersal, and the overall evolution of the gas PDF during $\tsfr$ is given by $\Msf/\taust + \Msf/\tm \pm \Msf/\tsfe$, and the associated timescale $(1/\taust+1/\tm \pm 1/\tsfe)^{-1}$ corresponds to the average time that gas spends in this state.  Thus, on average, gas will have to reach the star-forming state 
\begin{equation}
\label{eq:Nc} 
\Nc = \taust\left( \frac{1}{\taust} + \frac{1}{\tm} \pm \frac{1}{\tsfe} \right) = 1 + \xi + \frac{\taust}{\tmd} \pm \frac{\taust}{\tsfe}, 
\end{equation}
times, where we used Equation~(\ref{eq:Fm}) and the definition of the mass-loading factor, $\xi \equiv \taust/\tmfb$. 

Equation~(\ref{eq:tglob_xi}) is the key expression of our framework. It states that {\it the global depletion time is the sum of the total time that gas spends in the non-star-forming state over $\Nc$ cycles and the total time over which star-forming regions convert this gas into stars, $\taust$.} 

This equation elucidates how long $\tglob$ values can be reconciled with the relatively short local depletion times, $\taust$, and even shorter dynamical timescales, $\tp$, discussed in the Introduction. The global depletion time is longer than the depletion time in star-forming regions, $\taust$, due to the significant fraction of time that gas spends in the non-star-forming state. The global depletion time is longer than the timescale associated with dynamical processes supplying star-forming gas, $\tp$, because gas must evolve through the non-star-forming state $\Nc$ times, and $\Nc$ is large due to either efficient feedback, i.e., large $\xi$, or fast dynamical processes destroying star-forming regions, i.e., short $\tmd$ (see Equation \ref{eq:Nc}). 

When feedback dominates the removal of gas from the star-forming state, the number of cycles becomes $\Nc \sim \taust/\tmfb$. This clarifies how feedback can self-regulate star formation, i.e., how $\tglob$ can become insensitive to $\taust$. Indeed, the timescale $\tmfb$ is proportional to the rate of energy and momentum injection by feedback, which, in turn, is set by the local rate of star formation, i.e., $\taust$. Hence, $\tmfb\propto \taust$, which renders $\Nc$ insensitive to $\taust$. Thus, when $\Nc\tp\gg\taust$, the depletion time, $\tglob\approx \Nc\tp$, will be insensitive to $\taust$. 

In a nonequilibrium state, in which $\dot{M}_{\rm sf}>0$ ($<0$) during $\tsfr$, the term $\pm\taust/\tsfe$ in Equation~(\ref{eq:tglob_xi}) accounts for the correction of the average rates estimated using the star-forming gas fraction, $\fsf$, defined for the instantaneous masses $\Msf$ and $\Mg$. This correction appears because, when $\dot{M}_{\rm sf}>0$ ($<0$), the actual average fraction of $\tsfr$ that gas spends in the star-forming state is smaller (higher) than $\fsf$ and therefore more (fewer) transition cycles are required for depletion. 

In a steady state, on the other hand, the gas distribution is stationary and the star formation rate is in equilibrium with the gas fluxes into and out of the star-forming state: $\dot{M}_{\rm sf} = \Fsf - \SFR \approx 0$. In this case, $\tsfe\rightarrow\infty$ and the term $\taust/\tsfe$  can be neglected in Equation~(\ref{eq:tglob_xi}). In such a steady state, $\tglob = \Mg/\Fsf$ (see Equation~\ref{eq:tglob_Fsf}), and depletion time is determined by the \textit{net} rate of gas inflow into the star-forming state, $\Fsf$.  When $\Fsf$ is small, the depletion time is long. Galaxies as a whole reach the steady state with $\dot{M}_{\rm sf} \approx 0$ on the shortest of the timescales that control the global depletion time in Equation~(\ref{eq:tglob_xi}). Thus, globally, such an assumption is justified. However, individual ISM patches may deviate from the steady state, and the $\taust/\tsfe$ term will be one of the sources of the scatter in depletion times.  

In the remainder of the paper, we illustrate the framework described above using the results of an isolated galaxy simulation. Although the simulation adopts specific choices for many parameters, including resolution and prescriptions for star formation and feedback, the overall features and implications of our model do not depend on these specific choices. Our framework generically allows one to relate the depletion time on a large scale, e.g., the scale of an entire galaxy, to the star formation and feedback model that operates on a smaller scale, e.g., resolution scale of a simulation, where the distribution of local depletion times, $\tdep$, is defined.

\section{Simulation}
\label{sec:simulations}

To illustrate the framework outlined above and elucidate the physical processes that give rise to long global depletion times, we use a simulation of an isolated $\sim L_*$-sized galaxy. We carried out the simulation with the adaptive mesh refinement $N$-body and gas dynamics code ART \citep{Kravtsov.1999,Kravtsov.etal.2002,Rudd.etal.2008,Gnedin.Kravtsov.2011} and followed the evolution of an isolated gaseous disk in a live potential of a dark matter halo, stellar bulge, and stellar disk that are modeled with collisionless particles. 

We adopt the initial conditions that were used in the AGORA code comparison project \citep{agora2} and also in the studies of \citet{Agertz.etal.2013} and \citet{Semenov.etal.2016}. Specifically, the isolated disk is initialized inside a dark matter halo with $v_{\rm c,200} = 150 \kms$ and an initial concentration of $c = 10$. The initial disk of old stars has an exponential density profile with a radial scale length of $r_{\rm d} \approx 3.4 \kpc$ and a vertical scale height of $h_{\rm d} = 0.1 r_{\rm d}$ with a total mass of $M_{\rm \star,d} \approx 3.4 \times 10^{10} \Msun$. The stellar bulge has an initial mass of $M_{\rm \star,b} \approx 4.3 \times 10^9 \Msun$ that is distributed with a Hernquist density profile with $a = 0.1 r_{\rm d}$ \citep{Hernquist.1990}. The initial exponential gaseous disk has the same $r_{\rm d}$ and $h_{\rm d}$ as the stellar disk; its total mass is $\Mg \approx 8.6 \times 10^{9} \Msun$, which corresponds to the disk gas fraction of $\fgas \equiv \Mg/(M_{\rm \star,d}+\Mg) = 20\%$. 

Gas evolution is governed by modified hydrodynamical equations that include terms related to cooling and heating, dynamical effects of subgrid turbulence, gas consumption by star formation, and injection of mass, momentum, and energy by feedback from young stars.

Cooling in the optically thin limit is implemented following the model of \citet{Gnedin.Hollon.2012}. We assume a fixed metallicity of $Z = Z_\odot$ and constant background heating by interstellar radiation in the Lyman-Werner band with the photodissociation rate of $10^{-10}\ \rm s^{-1}$ \citep{Stecher.Williams.1967}. To model temperatures in dense self-shielded gas, we assume that extinction is proportional to the local column density of atomic gas, which we approximate as $n L_{\rm J,40}$, where $n$ is the gas number density in a cell and $L_{\rm J,40}$ is the local Jeans length with an applied temperature ceiling of $40\K$ \citep[model ``L1a'' in ][]{Safranek-Shrader.etal.2017}.

In our simulation, we adopt the unresolved turbulence model of \citet{Schmidt.etal.2014} that dynamically follows the kinetic energy of gas motions on subgrid scales, $K$, as a separate hydrodynamical field akin to thermal energy. Subgrid turbulence is generated by random velocities on the resolution scale; it exerts pressure and viscous forces on resolved gas motions and decays on the local cell-crossing timescale, as motivated by simulations of both subsonic and supersonic turbulence \citep[e.g., ][]{MacLow.etal.1998}. For details about the model implementation in ART, we refer the reader to \citet{Semenov.etal.2016}.

We adaptively resolve cells where the total gas mass exceeds $\sim 8\,300 \Msun$ and reach a maximum resolution of $\Delta = 40 \pc$. Such a $\Delta$ is sufficient to resolve ISM structure down to densities of $n \sim 100 - 1\,000 \cc$, and therefore we do resolve the dynamical processes that are sometimes claimed to limit the star-forming gas supply from the general ISM with average density of $n \sim 1\cc$. At the highest resolution level, we do not apply an artificial pressure floor in cold gas. Thus, the densities of star-forming regions are limited only by the effects of stellar feedback and the effective pressure due to thermal and both subgrid and resolved turbulent motions.

As a star formation prescription, we adopt the local rate $\rhoSFR=\rho/\tdep$, where $\rho$ is the gas density in a cell, and the local depletion time, $\tdep$, is related to the free-fall time, $\tff\equiv \sqrt{3\pi/32G\rho}$, with the efficiency per free-fall time, $\tdep = \tff/\epsff$. Numerical and analytical models of star formation in turbulent GMCs generally predict a strong increase of $\epsff$ with an increasing relative importance of gravity that facilitates star formation, in comparison to turbulence that provides support against gravity \citep[for a review, see][]{Padoan.etal.2014}. For example, \citet{Padoan.etal.2012} found that the star formation efficiency of a turbulent cloud increases exponentially with a decreasing virial parameter, $\epsff \approx \exp( - \sqrt{\avir/0.53} )$, where $\avir$ is defined for a box with a side $\Delta$ as for a uniform sphere of radius $R = \Delta/2$:

\begin{equation}
\label{eq:avir}
\avir \equiv \frac{5 \sigma_{\rm 1D}^2 R}{GM} \approx 9.35 \frac{ (\stot/10\kms)^2 }{ (n/100\cc) (\Delta/40 \pc)^2}. 
\end{equation}

In our simulations, to apply the \citet{Padoan.etal.2012} fit in thermally supported gas, in the definition of $\avir$ we consider contributions of both the sound speed, $\cs$, and the explicitly modeled subgrid turbulent velocities, $\st \equiv \sqrt{2 K/\rho}$:
\begin{equation}
\label{eq:sigma_tot}
\stot = \sqrt{ \st^2 + \cs^2 }. 
\end{equation}
Also, even though we are able to model $\epsff$ following the \citet{Padoan.etal.2012} formula \citep[see, e.g.,][]{Semenov.etal.2016,Li.etal.2017}, in the simulation used here, we approximate the continuous exponential dependence of $\epsff$ on $\avir$ assuming a constant $\epsff = 1\%$ for $\avir < \avirsf = 10$, and $\epsff=0$ elsewhere. This simulation was carried out as part of a simulation suite in which the $\epsff$ value was varied systematically to explore its effect on star formation in galaxies. Using a constant $\epsff$ and a sharp $\avirsf$ threshold makes the interpretation of simulation results easier, and we will present the results of this simulation suite in a forthcoming study. We explicitly checked that the global depletion times and the Kennicutt-Schmidt relations are similar in runs where $\epsff$ follows the \citet{Padoan.etal.2012} fit and where we approximate this fit with a threshold. A qualitatively similar star formation prescription but with a different choice of parameters was studied by \citet{Hopkins.etal.2013}.

We stress that the scenario of gas depletion described in Section~\ref{sec:model} remains valid for any choice of star formation prescription, although in Section~\ref{sec:discussion:simulations} we argue that such a prescription should be chosen carefully, as it is important for the prediction of realistic ISM properties. We note that our adopted threshold value, $\avirsf=10$, is consistent with the observed distribution of $\avir$ in molecular clouds with sizes comparable to our resolution of $\Delta = 40\pc$: such clouds tend to have $\avir \lesssim 10$ \citep{MivilleDeschenes.etal.2016}. The adopted value of $\epsff = 1\%$ is consistent with the average values deduced for observed actively star-forming clouds.  

In our feedback prescription, we inject momentum and energy from the type II supernovae (SNe) at a uniform rate between 3 and 43 Myr after the formation of a stellar particle. The total number of SNe exploded over this time interval is computed assuming the \citet{Chabrier.2003} IMF. The thermal energy and radial momentum injected by each supernova are calibrated against simulations of a SN remnant evolution in a nonuniform medium \citep{Martizzi.etal.2015}, taking into account ambient gas density and with an additional boost of radial momentum by a fiducial factor of 5. We adopt such a boosting factor to compensate for the numerical loss of injected momentum and to account for the actual physical uncertainties of the radial momentum estimates \citep[e.g.,][]{Gentry.etal.2017}. In addition to SNe type II feedback, during evolution, stellar particles return a fraction of their mass following the prescription of \citet{Leitner.Kravtsov.2011}.

\begin{figure*}
\centering
\includegraphics[width=\textwidth]{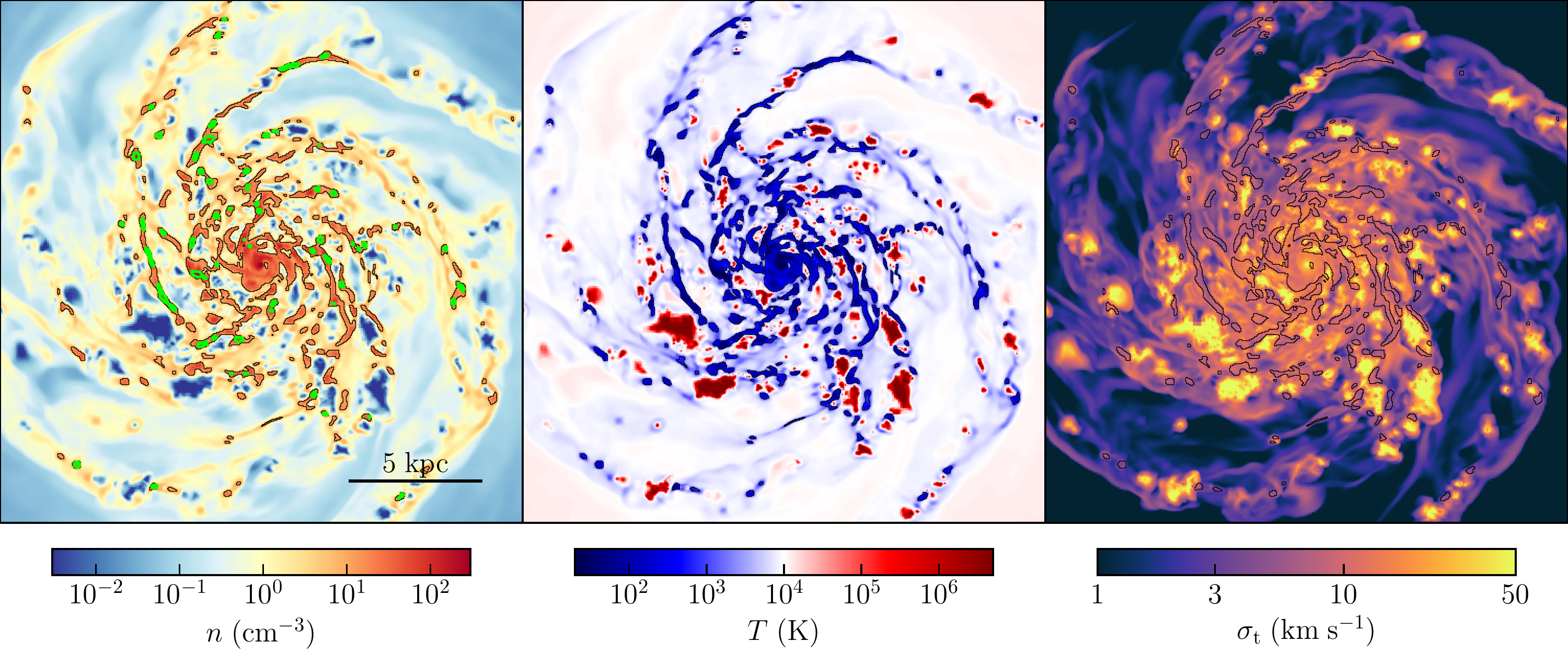}
\caption{\label{fig:faces} Midplane slices of gas number density, $n$, temperature, $T$, and subgrid turbulent velocity, $\st = \sqrt{2 K/\rho}$, after $500 \Myr$ of evolution. To make comparison easier, the black contours in all panels correspond to $n = 10 \cc$, above which the molecular mass fraction rapidly increases at solar metallicity. Green contours in the left panel indicate gas that satisfies our star formation criterion, $\avir < \avirsf = 10$. }
\end{figure*}

Figure~\ref{fig:faces} shows the spatial distribution of gas number density, temperature, and subgrid turbulent velocity in our simulated galaxy at $t = 500\Myr$. The figure highlights the multiphase, dynamic structure of the ISM. Comparison with a simulation with no feedback, which we will present in a companion paper, shows that the structure of the ISM is significantly affected by stellar feedback. Its effect is manifested not only in the ubiquitous regions of hot, turbulent gas but also in the overall flocculent nature of the spiral pattern.

In our analysis, we estimate the molecular gas density in each cell using the KMT model \citep{KMT1,KMT2,McKee.Krumholz.2010}: $\rhoH2 = \max[0,(1-0.75 s/(1+0.25 s))\rho]$, where at solar metallicity $s\approx 1.8/\tau_{\rm c}$ and $\tau_{\rm c} = 320 (\rho \Delta / {\rm g\ cm^{-2}})$. This model predicts a rapid increase of the molecular fraction at densities $n>10 \cc$ indicated by the black contour in Figure~\ref{fig:faces}. The resulting total mass fraction of molecular gas in our simulation is $\fH2 \equiv \MH2/\Mg \sim 20\%$.

Subgrid turbulent velocities, $\st$, range from $\lesssim 3 \kms$ in the diffuse ISM between the spiral arms to $\sim 30-300 \kms$ in hot SNe bubbles. In this simulation, supernovae do not explicitly inject turbulent energy, and high $\st$ in hot bubbles are generated by the subgrid turbulence model. In the cold dense gas, turbulent velocities are supersonic and also vary significantly, $\st \sim 5 - 15 \kms$. Strong subgrid turbulence in cold gas results in high values of $\avir$ and, according to our star formation criterion, $\avir < \avirsf = 10$, only $\sim 40 \%$ of all molecular gas mass is star-forming at any given moment. Such star-forming regions are shown in the left panel of Figure~\ref{fig:faces} with green contours.

The total SFR of our model galaxy is $\SFR \sim 1-2 \Msunyr$, which translates to global depletion times for the total and molecular gas of $\tglob \sim 4-8 \Gyr$ and $\tH2 \sim 1-2 \Gyr$. The values of the depletion times and molecular fraction are in the ballpark of the typical values observed in nearby spiral galaxies \citep[e.g.,][]{wong_blitz02,Bigiel.etal.2008,Leroy.etal.2013}. 

\begin{figure*}
\centering
\includegraphics[width=\textwidth]{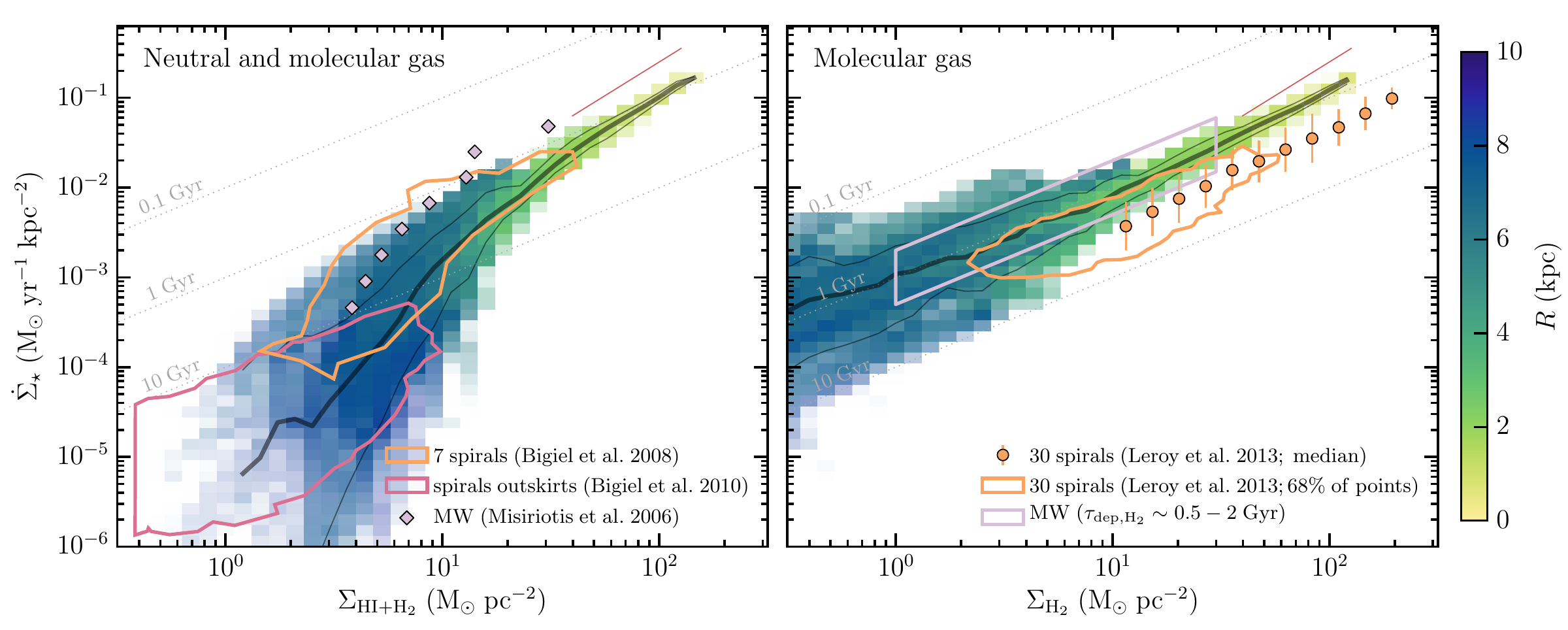}
\caption{\label{fig:KS} Relation between the surface density of the SFR and the total (left panel) and molecular gas (right panel) in our simulation and in observations. To match the typical spatial scales on which this relation is usually measured, we smooth 2D maps of $\SSFR$, $\Sigma_{\rm HI+H_2}$ and $\SH2$ obtained at 500 Myr with a Gaussian filter with a width of 1 kpc and plot the median, 16$^{\rm th}$, and 84$^{\rm th}$ percentiles of the resulting pixel distribution (thick and thin gray lines). The distributions are colored according to the average galactic radius, $R$, of pixels in a given bin. To match the averaging timescales of the star formation indicators, we measure the surface density of stars that are younger than $30\Myr$ and define $\SSFR$ as $\Sigma_\star(<30 \Myr)/30 \Myr$. The thin red line at high $\Sigma$ indicates the slope adopted in our star formation prescription, $\rhoSFR \propto \rho^{1.5}$. Thin dotted lines correspond to the constant depletion times of 0.1, 1, and 10 Gyr (from top to bottom). We compare our results to the observed relations in nearby spiral galaxies \citep{Bigiel.etal.2008,Bigiel.etal.2010,Leroy.etal.2013} and in the Milky Way \citep{Misiriotis.etal.2006}. The violet contour in the right panel shows the range of $\tH2 \sim 0.5 - 2 \Gyr$ estimated using the radial profiles of $\SSFR$ and $\SH2$ for the Milky Way from Figure 7 in \citet{Kennicutt.Evans.2012}. }
\end{figure*}

Figure~\ref{fig:KS} compares the Kennicutt-Schmidt relation between the surface densities of the SFR and \mbox{\ion{H}{1} + H$_2$} and H$_2$ gas in our simulation to the observed relations in the Milky Way and nearby spiral galaxies. Our results are in a good agreement in both normalization (i.e., the depletion time value) and slope. Note, in particular, that the linear relation between $\dot\Sigma_\star$ and $\Sigma_{\rm H_2}$ emerges from the nonlinear star formation prescription adopted in our simulation: $\rhoSFR \propto \rho^{1.5}$. In Section \ref{sec:results:KS}, we consider the origin of the linear relation and show that it results from the particular behavior of the gas density distribution shaped by stellar feedback. 

For a consistent comparison with the observed KSR for \mbox{\ion{H}{1} + H$_2$} gas, in our simulation, we defined neutral hydrogen to be all nonmolecular gas denser than $n_{\rm H,SSh}$, given by Equation (13) in \citet{Rahmati.etal.2013}. This threshold corresponds to the gas self-shielded from the far ultraviolet (FUV) background with the adopted photoionization rate $\Gamma = 10^{-10}\ \rm s^{-1}$. We also excluded all neutral hydrogen that is colder than $1000 \K$ assuming that it constitutes the optically thick cold neutral medium (CNM) not included into the observed measurements of $\Sigma_{\rm HI}$. Our temperature threshold is somewhat higher than the CNM temperature estimated in real galaxies \citep[$\lesssim 300\K$, e.g.,][]{Wolfire.etal.2003} because, in our simulation, we do not resolve the transition between warm and cold neutral gas phases, which results in intermediate gas temperatures on the resolution scale. The particular value of the temperature threshold was chosen to select $\sim 40\%$ of the neutral hydrogen mass, which is close to the CNM mass fractions estimated in the Milky Way and nearby galaxies \citep[e.g.,][]{Heiles.Troland.2003,Braun.2012,Pineda.etal.2013,Sofue.2017}.

In the analyses presented below, we consider the processes in the ISM during the time interval between $400$ and $600 \Myr$, which is short enough to neglect the effect of global gas consumption on the ISM dynamics and long enough to accumulate sufficient statistics of rare events and make the considered distributions representative and well-sampled. 

To study the detailed dynamics of individual gas parcels, we use gas tracer particles that are passively advected with the local flow velocity interpolated to the positions of the particles with the cloud-in-cell scheme. We populate the disk with $10^5$ tracer particles uniformly initialized within $R < 8 \kpc$ after $300 \Myr$ of disk evolution when the transients related to the initial off-equilibrium state had dissipated away. After initialization, we wait for $100 \Myr$ to let the tracers equilibrate with the gas density distribution. At that point, the distributions of tracer densities in radial annuli resemble the gas density PDF in computational cells, and the distribution of tracer properties thus can be considered to be a good approximation of the mass-weighted PDF of gas properties. 

We average the distribution of tracers to construct statistics, such as their PDF and fluxes in the $n$~--~$\stot$ phase diagram, between $400$ and $600 \Myr$ with $1 \Myr$ steps. We checked that at every moment between $400$ and $600 \Myr$ phase distributions of gas and tracer particles resembled their averaged versions, which means that the galaxy remained in approximate equilibrium  over the considered period of time.

We focus on the evolution of gas in the phase plane of density, $n$, and total velocity dispersion, $\stot$, because the position of a gas parcel in this plane determines its internal consumption time, $\tdep$, according to our star formation prescription. To quantify gas motions in the $n$~--~$\stot$ plane, we measure the derivatives $d \log n / dt$ and $d \log \stot / dt$ probed by each tracer every $1 \Myr$. To estimate the average local flow rates of gas, we accumulate fluxes corresponding to these derivatives and normalize them by the local density of tracers in the $n$~--~$\stot$ plane.

To characterize actual fluxes that supply and remove star-forming gas, in addition to the total flux of tracers, we separately track the fluxes of tracers with decreasing or increasing $\avir$. We quantify the magnitudes of these fluxes with the characteristic evolution timescale, on which $\avir$ changes by an order of magnitude at a given rate,
\begin{equation}
\label{eq:tau_alpha}
\tau_{\alpha,\gtrless 0} \equiv \left\langle \left| \frac{d \log_{10} \avir}{dt} \right| \right\rangle^{-1}_{\gtrless 0},
\end{equation}
where we average the derivative of $\log_{10} \avir$ taking into account only tracers with decreasing (increasing) $\avir$ to compute $\tau_{\alpha,< 0}$ ($\tau_{\alpha,> 0}$).

\section{Results}
\label{sec:results}

\subsection{The origin of long global depletion times}
\label{sec:results:phases}

\begin{figure}
\centering
\includegraphics[width=\columnwidth]{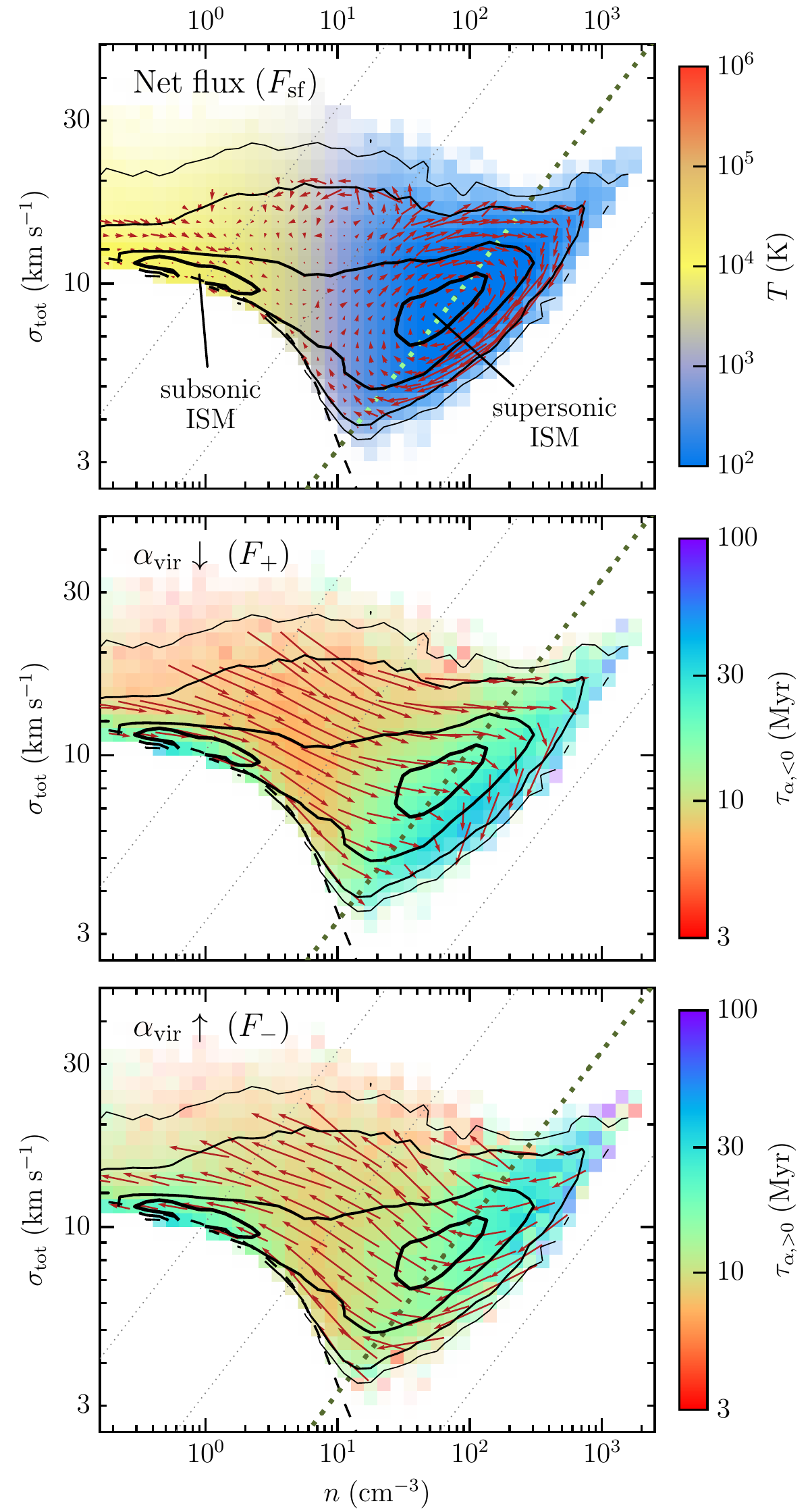}
\caption{\label{fig:phases_fid} Distribution of gas tracer particles in the plane of gas number density, $n$, and total velocity dispersion, $\stot = \sqrt{\st^2+\cs^2}$, averaged between $400$ and $600 \Myr$. The black contours in all panels indicate the average PDF of tracers and correspond to 20, 68, 95, and 99\% of all tracers. The diagonal dotted lines indicate constant values of $\avir$ from left to right: 1000, 100, 10, and 1, with $\avir = \avirsf = 10$ shown by the thick green dotted line. The dashed line along the lower envelope of the PDF at $n<10\cc$ indicates the median sound speed, $\cs$, in each density bin. Colors in the top panel show the average gas temperature in each bin, while arrows indicate the average total fluxes of gas tracers measured as described at the end of Section~\ref{sec:simulations}. Arrows in the middle and bottom panels correspond to the fluxes of gas tracers with decreasing and increasing $\avir$ respectively. Colors in these panels show the distribution of the characteristic evolution timescales, $\tau_{\alpha,\gtrless 0}$, defined by Equation~(\ref{eq:tau_alpha}). The normalizations of the arrows are the same in all three panels and correspond to the distances that tracers would traverse at a given rate over $5\Myr$. }
\end{figure}

As we discussed at the end of Section~\ref{sec:model}, in steady state, when $\langle\dot{M}_{\rm sf}\rangle_{\tsfr} \approx 0$ on the timescale $\tsfr$ over which the SFR is estimated, long global depletion times, $\tglob \equiv \Mg/\SFR = \Mg/\Fsf$, originate from a small \textit{net} flux of gas into the star-forming state, $\Fsf$. In principle, $\Fsf$ could be small if the rate at which gas evolves toward the star-forming state were set by a slow ``bottleneck'' process. However, as Figure~\ref{fig:phases_fid} shows, in simulations with efficient feedback, gas rapidly transitions between the star-forming and non-star-forming states, and a small $\Fsf$ results from a near-cancellation of large opposite fluxes into and out of the star-forming state.

In this figure, we plot the distribution of gas tracer particles within the disk in the plane of gas number density, $n$, and $\stot = \sqrt{\st^2+\cs^2}$, that can be viewed as an effective temperature including both thermal and turbulent gas motions on subgrid scales. The gas distribution spans a wide range of densities, $\stot$, and temperatures and has two distinct peaks. The peak at low densities, $n \sim 1 \cc$, corresponds to diffuse, warm, subsonic ($\st \lesssim \cs$) gas at temperature $T \sim 10^4 \K$. The gas in the second peak at $n > 10 \cc$, on the other hand, is cold ($T\lesssim 100$ K) and supersonic ($\st>\cs$). 

According to our star formation prescription, the star-forming gas has $\avir<\avirsf=10$. Such gas in Figure~\ref{fig:phases_fid} resides below the thick green dotted line. The net mass flux of gas in the $n-\stot$ plane is visualized by the arrows in the top panel of Figure~\ref{fig:phases_fid}, where the length of the arrows is equal to the distance tracers would traverse in $5\Myr$ for a given flux. The figure shows that arrows are rather small throughout most of the phase space occupied by tracers and are particularly small near the thick green dotted line. This means that the \textit{net} evolution of gas in the $n$~--~$\stot$ plane is slow and the net flux through the star formation threshold, $\Fsf$, is small. This small net flux results in the long global depletion timescales exhibited by our simulated galaxy, $\tglob \sim 5 \Gyr$ and $\tH2 \sim 1 \Gyr$ (see Section~\ref{sec:simulations} and Figure~\ref{fig:KS}).

However, the middle and bottom panels of Figure~\ref{fig:phases_fid} show that the small net $\Fsf$ results from the near-cancellation of two opposite fluxes. These panels show the fluxes of only those tracers in which $\avir$ is decreasing, $\Fp$, or increasing, $\Fm$, and these fluxes are significantly stronger than the net flux in the top panel. A typical tracer evolves toward and away from the star-forming state on a timescale of order $\tau_{\alpha,\gtrless 0} \sim 5-30$ Myr, consistent with the estimates of the timescales of relevant processes in Section~\ref{sec:introduction}. Thus, the rate of gas supply from the diffuse, warm ISM to the star-forming state cannot be the factor limiting the global star formation rate, as envisioned by \citet[][]{Saitoh.etal.2008}. Instead, gas generally evolves from the diffuse to the star-forming state on a timescale of tens of Myr, much shorter than the global depletion time. The latter is long because gas rapidly leaves the star-forming state at the rate that nearly cancels the rate at which gas is reaching this state. In the next section, we consider the processes that drive the fast gas evolution in more detail.

\vspace{1em}
\subsection{Dynamical processes shaping ISM}
\label{sec:results:paths}

\begin{figure*}
\centering
\includegraphics[width=\textwidth]{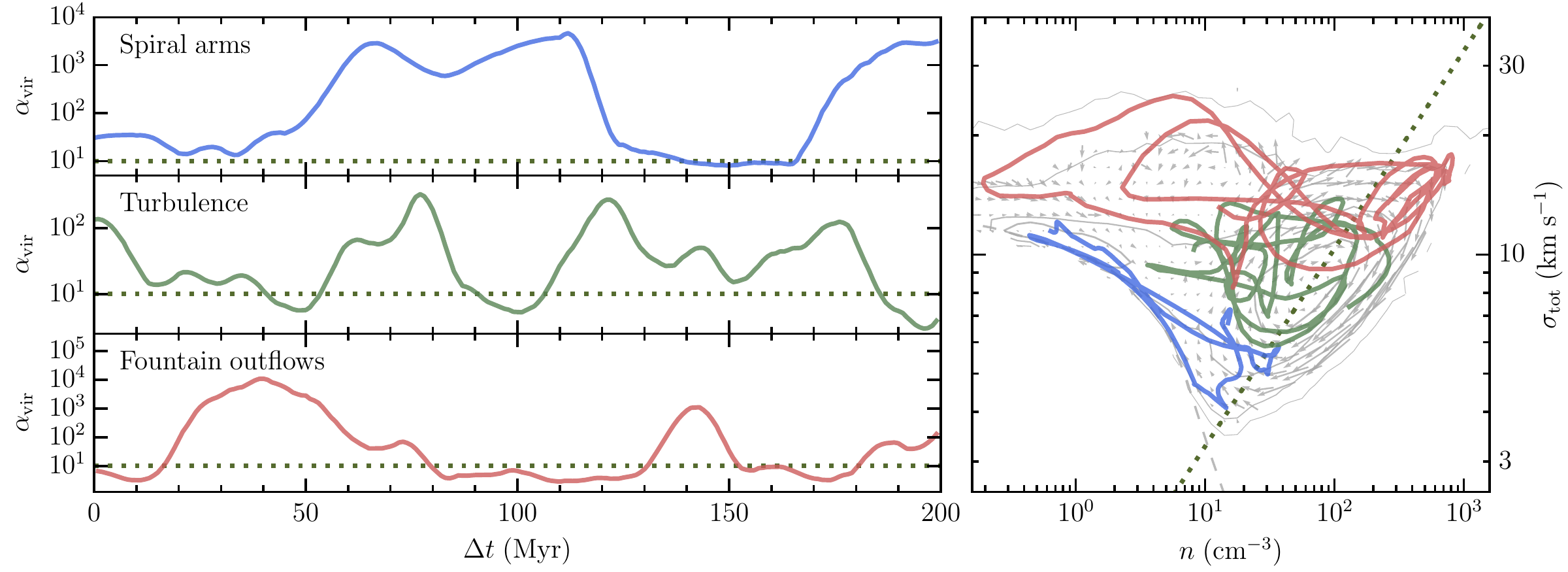}
\caption{\label{fig:paths} Trajectories of three illustrative tracers followed for $200 \Myr$. The left set of panels shows the evolution of $\avir$ for each of the three tracers. The right panel shows the trajectories in the $n$~--~$\stot$ plane with the corresponding colors. The gray contours and arrows indicate the average tracer PDF and their net fluxes, as in the top panel of Figure~\ref{fig:phases_fid}. The thick dotted lines in all panels correspond to the adopted star formation threshold, $\avir = \avirsf = 10$. For presentation purposes, small fluctuations of actual tracer trajectories on timescales $\lesssim 5 \Myr$ were smoothed using Savitzky-Golay filter. }
\end{figure*}

\begin{figure}
\centering
\includegraphics[width=\columnwidth]{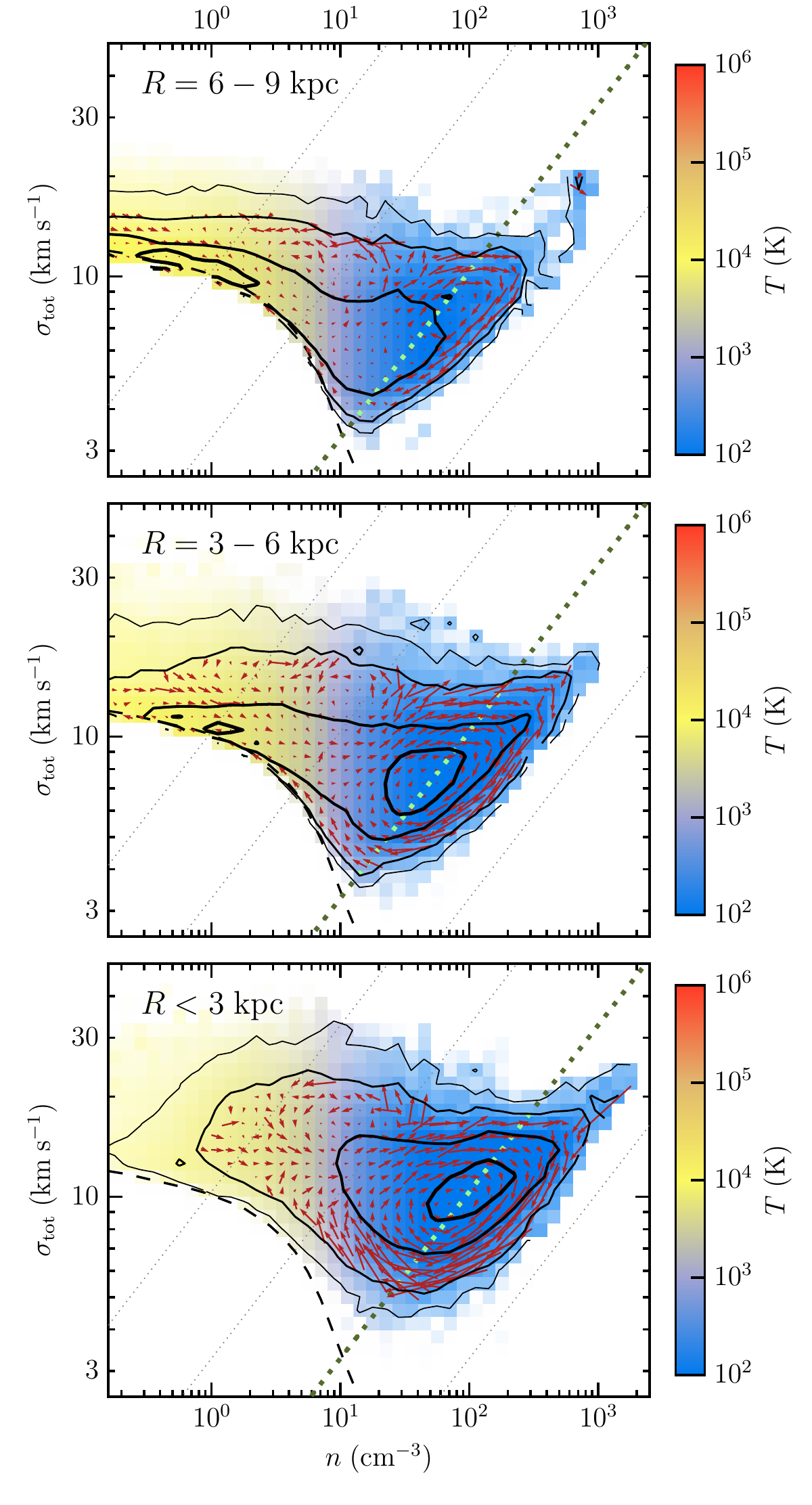}
\caption{\label{fig:phases_r} Distributions and average fluxes of tracers residing at different galactic radii, $R$ (shown in the top left corner of each panel). Notation follows that of the top panel of Figure~\ref{fig:phases_fid}. Comparison with the right panel of Figure~\ref{fig:paths} hints that the distribution of tracers on the disk outskirts (top panel) is predominantly shaped by gas compression and expansion due to the spiral arms, while close to the disk center (bottom panel), the distribution is shaped by feedback-driven turbulence and outflows. Whirl-like patterns of velocities in the cold, supersonic phase indicate that the distribution of dense gas at all radii is affected by star formation feedback (see text for details). }
\end{figure}

The average gas flow patterns shown with arrows in Figure~\ref{fig:phases_fid} result from the statistical averaging of the complicated trajectories of individual tracer particles. The particular shapes of such trajectories vary depending on local conditions and specific physical processes that govern gas evolution.

In our simulation, gas evolution between diffuse, warm, subsonic and dense, cold, supersonic ISM phases is governed by large-scale disk instabilities and the turbulent flows generated by them. The evolution of gas in the dense phase is determined predominantly by stellar feedback that disperses star-forming regions, drives large-scale ISM turbulence, and launches fountain-like outflows. 

In the following subsections, we consider these processes using three illustrative tracer trajectories integrated over $200 \Myr$ and shown in Figure~\ref{fig:paths}. We chose these particular tracers because their evolution over the considered period of time is governed predominantly by the same process over several consequential cycles of compression and expansion.

\subsubsection{Compression and expansion due to spiral arms}
\label{sec:paths:spiral-arm}

The blue line in the top left panel of Figure~\ref{fig:paths} shows an example of the $\avir$ evolution followed by a tracer that swings between the subsonic and supersonic phases during cycles of compression and expansion as it enters and exits the spiral arms. The trajectory of this tracer in the $n$~--~$\stot$ plane is shown with the same color in the right panel.

Due to strong compression, gas entering a spiral arm rapidly cools down and loses the thermal support that initially dominates in the subsonic ISM. At the same time, initially low subgrid turbulent velocities of the subsonic ISM, $\st \lesssim 3 \kms$, rapidly grow due to compressional heating (see the detailed discussion in Section~3.1 of \citealt{Semenov.etal.2016} and also \citealt{Robertson.Goldreich.2012}). At $n \sim 10 \cc$, when subgrid turbulent velocities become comparable to the thermal speed, gas detaches from the lower envelope of the distribution shown in Figure \ref{fig:phases_fid} and enters the supersonic ISM phase. Similarly, when gas leaves a spiral arm, it expands, and subgrid turbulent velocities decrease. Eventually, under the influence of expansion and interstellar FUV heating, gas returns to the subsonic ISM phase with $n \sim 1 \cc$ and $T \sim 10^4 \K$. 

The actual transition of gas between the subsonic and supersonic phases is fast, as it is controlled by a strong compression rate in the spiral arms and short cooling times at $n > 1\cc$. Hence, the rate at which diffuse gas is promoted into the dense phase is mostly determined by the time that gas waits between subsequent passages of the spiral arms, 

\begin{equation}
\tau_{\rm arm} \sim \frac{2 \pi R}{m V_{\rm gas}} \sim 80\Myr \frac{ (R/8 \kpc) }{ (m/6)(V_{\rm gas} / 100 \kms)},
\end{equation}
where $V_{\rm gas} \equiv v_{\rm gas}-v_{\rm pat}$ is the speed of gas relative to the spiral waves pattern, and we set $m=6$, as our simulated galaxy develops six spiral arms. 

The typical time that gas spends inside a spiral arm before expansion contributes to the dynamical rate of gas removal from the star-forming state, $\tmd$. This timescale depends on the spiral arm width, gas velocity, and the angle at which gas flows inside the arm. Depending on local conditions, this timescale can be as long as a few tens of Myr.

In Figure~\ref{fig:phases_r} we plot distributions of $n$ and $\stot$ separately for tracers residing at different galactic radii and therefore experiencing different ISM conditions. The distribution in the outer disk (top panel) is shaped predominantly by the compression and expansion due to the spiral arms. Specifically, most of the gas mass in the outer disk resides in the diffuse subsonic phase and forms a peak at $n \sim 1 \cc$ and $T \sim 10^4 \K$. The tail extending along the lower envelope of the distribution toward the dense supersonic phase corresponds to the gas currently being compressed in the spiral arms. As the figure also shows, the compression of diffuse gas in the spiral arms is only relevant at large radii, whereas closer to the disk center, less gas remains in the diffuse phase, and this process becomes much less important.

\subsubsection{SNe-induced shocks and ISM turbulence}
\label{sec:paths:turbulence}

We find that the evolution of dense, supersonic gas in the $n$~--~$\stot$ plane is dominated by the turbulence that is driven by stellar feedback. Injection of momentum by SNe in a star-forming region results in a rapid expansion of gas until the region is eventually dispersed. Shocks associated with expanding bubbles compress gas in the disk plane, which may induce new episodes of star formation and subsequent SN explosions. The turbulence resulting from overlapping and interacting bubbles makes gas parcels oscillate in fast cycles, as illustrated by the green trajectory in Figure~\ref{fig:paths}.

The characteristic timescale between subsequent compressions of ISM gas by such expanding SN shocks corresponds to
\begin{equation}
\tau_{\rm shell} \sim \frac{L}{v_{\rm shell}} \sim 50\Myr \frac{ ( L/1 \kpc) }{ ( v_{\rm shell}/20 \kms)},
\end{equation}
where $L$ is a typical separation between bubbles (see, e.g., the temperature map in the middle panel of Figure~\ref{fig:faces}) and $v_{\rm shell}$ is a typical velocity of shells on a scale $L$.  

Compression and expansion of gas in the turbulent ISM is accompanied by the increase and decrease of turbulent velocity dispersion. As a result, averaging of such large-scale turbulent motions over many tracers results in a prominent clockwise whirl-like pattern of arrows around the peak of the PDF in the cold (blue) part of the diagram (see Figures~\ref{fig:phases_fid} and \ref{fig:phases_r}). Closer to the peak center, the net flux magnitude decreases due to the averaging between fast motions of many tracers at the different stages of their turbulent compression-expansion cycles.

As Figure~\ref{fig:phases_r} shows, such a whirlwind pattern is most prominent at $R\leq 6$ kpc. Thus, the feedback-driven turbulence and associated compression and expansion of gas are dominant processes at these radii in the cold, supersonic gas. The ISM at these radii has a complex structure (see Figure~\ref{fig:faces}) reflecting the chaotic turbulent nature of the gas.

\subsubsection{Feedback-driven fountain outflows}
\label{sec:paths:fountains}

Supernova feedback also affects some of the gas by accelerating it in the direction perpendicular to the disk plane. Such gas expands in fountain-like outflows but eventually cycles back to the ISM under the influence of the disk potential. Interactions of such outflows with the halo gas adjacent to the disk result in an increase of small-scale turbulent velocities that quickly dissipate when the gas falls back onto the disk. 

An example of a tracer trajectory during expansion and subsequent recycling of a fountain outflow is shown with the red lines in Figure~\ref{fig:paths}. This particular tracer was ejected and recycled twice, at $\sim 20-80$ and $\sim 130-150 \Myr$. In each event, after its star-forming region was dispersed by feedback, this tracer acquired a moderate vertical velocity of $v_z \sim 50 \kms$ and elevated as high as $\sim 400 \pc$ above the disk plane, i.e., a few scale heights, before falling back onto the disk. At the highest elevation point, the gas in these outflows expands only to the densities comparable to those of the diffuse subsonic ISM phase, $n \sim 0.2 - 2 \cc$ and its virial parameter reaches the values of $\avir \sim 10^3-10^4$ due to the strong turbulence generated by the interaction of the expanding outflow with the surrounding gas. Outflows launched by feedback from regions of more vigorous star formation reach even lower $n$ and higher $\avir$.

The timescale of the fountain cycle can be estimated as a dynamical time in the gravitational field of a massive infinite sheet of constant surface density $\Sigma_{\rm tot}$, corresponding to the local total surface density of the disk,
\begin{equation}
\tau_{\rm grav} \sim \frac{v_z}{\pi G \Sigma_{\rm tot}} \sim 20\Myr \frac{ (v_z/50 \kms) }{ (\Sigma_{\rm tot}/200 \Msun \pc^{-2})},
\end{equation}
where $v_z$ is the initial vertical velocity of gas in the outflow and $\Sigma_{\rm tot} = \Sigma_\star + \Sg \sim 200 \Msun \pc^{-2}$ is the typical total surface density of gas and stars in our simulated galaxies.

The averaging of trajectories between many gas parcels constituting fountain-like outflows results in a tail of the distribution directed from the star-forming state toward the lower densities and higher $\stot$. The total flux of tracers forms a prominent counterclockwise vortex inside this tail that is clearly seen in Figures~\ref{fig:phases_fid} and \ref{fig:phases_r}. 

Figure~\ref{fig:phases_r} shows that at all radii within the disk, some fraction of gas evolves in the manner discussed above, which indicates the existence of fountain-like outflows. At larger radii, where the SFR is slower, the outflows are less prominent but still visible as a net flux of tracers directed toward lower densities along the top envelope of the distribution shown in the upper panel. Gas in such outflows at large radii usually returns to the diffuse, warm, subsonic ISM between the spiral arms. Closer to the center, outflows are ubiquitous, and, after falling back, their gas directly rejoins the tumultuous large-scale turbulent motions of dense, supersonic gas.

\subsection{Duration and number of evolution cycles}
\label{sec:results:tc}

\begin{figure}
\centering
\includegraphics[width=\columnwidth]{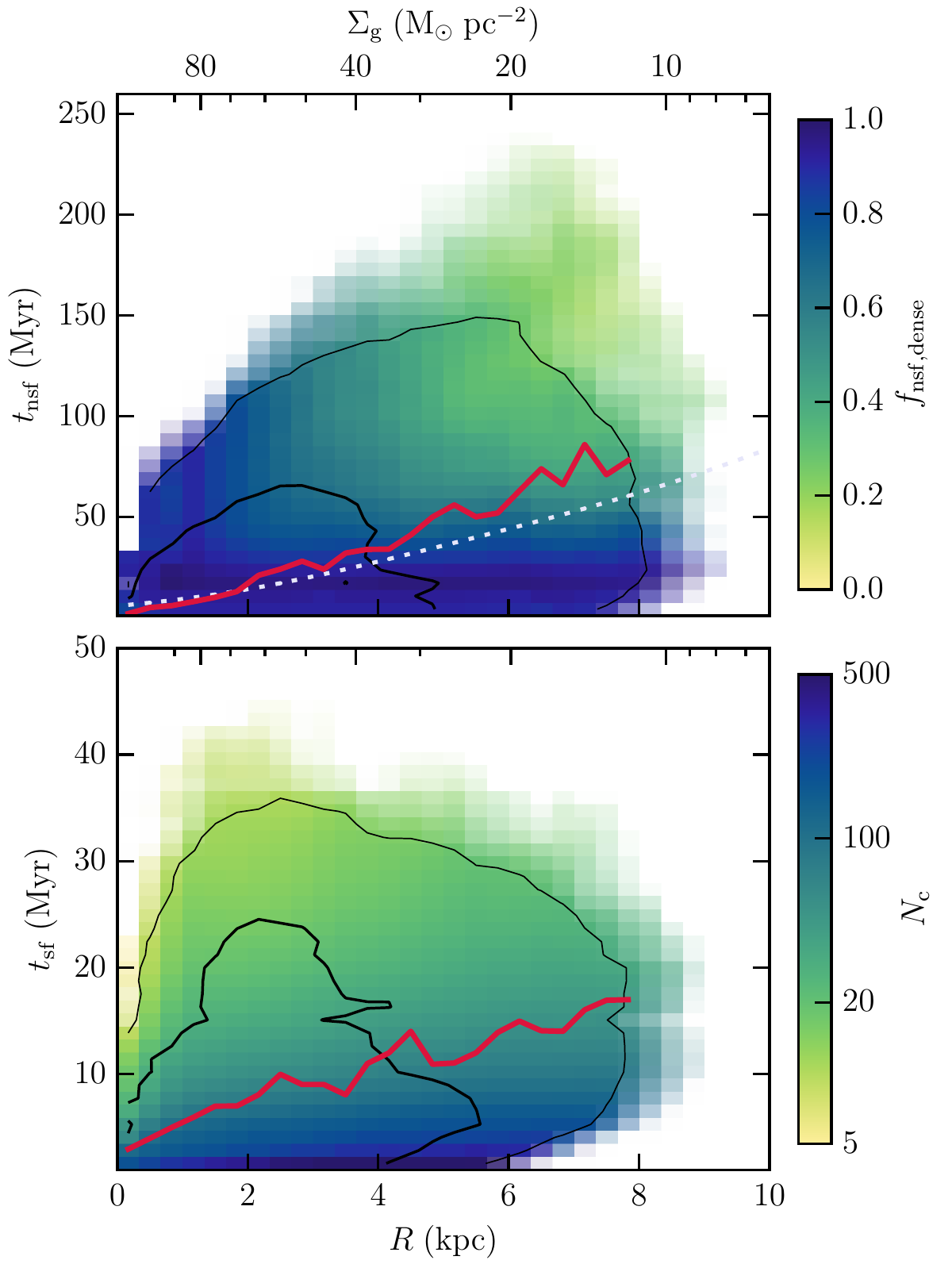}
\caption{\label{fig:tcdist} Distribution of time that tracers spend in non-star-forming ($\tnsf$, top panel) and star-forming ($\tsf$, bottom panel) states between successive crossings of the star formation threshold. Distributions of $\tnsf$ and $\tsf$ are shown as a function of galactic radius, $R$. The top axis in each panel also indicates the average surface density of gas at a given radius, $\Sg \propto \exp(-R/r_{\rm d})$, where $r_{\rm d}$ is the initial scale radius of the disk. Contours indicate 68\% and 95\% of trajectories. The thick red lines show the median timescales at every radius. The dotted line in the top panel corresponds to the free-fall time at the average midplane density at a given radius. Colors in the top panel show the average fraction of time spent in the dense phase with $n > 10\cc$: $f_{\rm nsf,dense} \equiv t_{\rm nsf,dense}/t_{\rm nsf}$. Colors in the bottom panel show the average number of passages through the star-forming state required for complete depletion, $\Nc$ (Equation~\ref{eq:Nc_sf}). To increase the statistics for long cycles, we follow tracers between $400$ and $1\,000 \Myr$ of the disk evolution. For presentation purposes, we smooth the resulting distributions, preserving their main features. }
\end{figure}

Typical tracer trajectories considered in the previous section explicitly confirm that during the evolution, gas parcels perform many fast cycles and rapidly explore a significant portion of the PDF extent, frequently switching between non-star-forming and star-forming states. As we discussed in Section~\ref{sec:model}, the distribution of the time that the gas parcels spend on each cycle in these states, $\tnsf$ and $\tsf$, determines the global depletion time of the galaxy.

In Figure~\ref{fig:tcdist} we plot the distribution of $\tnsf$ and $\tsf$ directly measured from the trajectories of all tracers as the time between consequential crossings of the star formation threshold. The results in the previous section indicate that the mix of the processes governing gas evolution may change with the galactic radius, $R$, and surface density, $\Sg$. Thus, to explore possible trends, we plot the distributions of timescales as a function of $R$ and $\Sg$.

The distribution of $\tnsf$ shown in the top panel indicates that the majority of tracers spend less than $100 \Myr$ in the non-star-forming stage of evolution during each cycle.  At higher average surface densities closer to the disk center, this time is even shorter, $\tnsf \lesssim 50 \Myr$, with a very low median value (thick red line). 

Colors in the top panel show the average fraction of time that gas tracers spend in the dense phase, $n > 10\cc$, over the non-star-forming stage of evolution. A blue color at small radii implies that gas preferentially stays in the dense, molecular phase even when it does not form stars. This is also evident from the bottom panel of Figure~\ref{fig:phases_r}, which shows that only a small fraction of gas expands to $n < 10\cc$ and it does so as a part of fountain outflows.

At larger radii, the relatively slow rate of star-forming gas replenishment via compression in the spiral arms becomes important, and the median $\tnsf$ increases to $\sim 80 \Myr$. Gas governed by this process spends significant time in the diffuse subsonic ISM, and such tracer trajectories occupy the areas of the longest $\tnsf$ at $R > 4$ kpc (green color in Figure~\ref{fig:tcdist}). However, as indicated by the blue color, at such radii many tracers still perform short cycles with $\tnsf < 50 \Myr$ without leaving the dense phase.   
 
The increase of the $\tnsf$ median value is consistent with the scaling proportional to the free-fall time at the mean or midplane density at a given radius, $t_{\rm ff,0}\propto \rho_0^{-1/2} = (\Sg/2h_{\rm d})^{-1/2}$, shown by the dotted line in Figure~\ref{fig:tcdist} \citep[see also][]{Saitoh.etal.2008}. Such scaling is sometimes adopted in analytical models of galactic star formation to define the timescale on which star-forming regions are created \citep[e.g.,][]{Krumholz.etal.2012,Elmegreen.2015}. As we discussed above, $\tnsf$ in our simulations is set by both stellar feedback that drives turbulence and dynamical processes within the ISM. The scaling of the median $\tnsf$ with density indicates that gravity and the associated timescale plays at least some role in setting the time that gas spends in the non-star-forming state. For example, the fall of the gas driven out in a fountain outflow back to the disk will occur on a timescale of order $\sim t_{\rm ff,0}$. 

The bottom panel of Figure~\ref{fig:tcdist} shows the distribution of time spent by tracers in the star-forming state on each cycle. This timescale is close to the typical ``lifetime'' of star-forming regions and is quite short: $\tsf \lesssim 20 \Myr$ or $2-4$ free-fall times at the typical densities of star-forming regions. The fact that $\tsf$ is, on average, significantly shorter than $\tnsf$ is consistent with the small mass fraction of star-forming gas.

As we discussed in Section~\ref{sec:model}, the average time that a gas parcel spends in the star-forming state on a single cycle determines the total number of such cycles required for complete depletion as $\Nc = \taust/\tsf$. For every tracer on each passage through the star-forming stage, we estimate this number as the inverse fraction of mass depleted during the passage,
\begin{equation}
\label{eq:Nc_sf}
\Nc^{-1} = \int \frac{dt}{\tdep} = \int \epsff \frac{dt}{\tff},
\end{equation} 
where the integral is accumulated for each tracer particle while it is in the star-forming state between subsequent crossings of the star formation threshold. The resulting distribution of $\Nc$ is shown by the colors in the bottom panel of the figure. 

In agreement with our model, typical $\Nc \sim 50$ and the lifetimes of gas in the star-forming state, $\tsf \sim 10-20 \Myr$, are consistent with the range of the star-forming gas depletion times, $\taust \sim \Nc \tsf \sim 300-500$ Myr, obtained in our simulation (see the next subsection). In addition, assuming $\tnsf \sim 50-100 \Myr$, Equation~(\ref{eq:tglob1}) for the typical depletion time of a gas parcel gives a value of $\sim 2-5 \Gyr$, which is consistent with the actual global depletion time obtained in our simulation (see Section~\ref{sec:simulations}).

\subsection{Emergence of linear $\dot{\Sigma}_\star-\Sigma_{\rm H_2}$ relation}
\label{sec:results:KS}

One of the most intriguing results of our simulation is the emergence of the linear $\SSFR - \SH2$ relation consistent with observations (see the right panel of Figure~\ref{fig:KS}), even though on small scales the star formation rate is assumed to scale nonlinearly with the gas density:  $\rhoSFR = \epsff \rho/\tff \propto \rho^{1.5}$. This result provides a counterexample to the arguments that in simulations, the slope of the Kennicutt-Schmidt relation on large and small scales should be the same \citep{Schaye.DallaVecchia.2008,Gnedin.etal.2014}.

The observed linearity of the molecular KSR is often explained by the ``counting argument,'' which posits that all molecular gas is in star-forming clouds of approximately the same depletion time. In this case, the surface density of the SFR will vary linearly with the surface density of molecular gas, as long as the geometric covering fraction of star-forming clouds is less than unity. 

However, if not all of the molecular gas is star-forming, as is the case in our simulations and is likely the case in observed galaxies, the explanation for the linearity must be more nuanced. Analogously to Equation~(\ref{eq:tglob_def}), the depletion time of molecular hydrogen is given by
\begin{equation}
\label{eq:KS_tdep_H2}
\tH2 \equiv \frac{\SH2}{\SSFR} = \frac{\taust}{f_{\rm sf,H_2}},
\end{equation}
where $f_{\rm sf,H_2} \equiv \Ssf/\SH2$. The linearity of the molecular KSR, or, equivalently, the independence of the $\tH2$ of $\SH2$, must therefore arise from a lack of dependence of $\taust$ and $f_{\rm sf,H_2}$ on $\SH2$ or a cancellation of any such dependence in their ratio. 

The depletion time of star-forming gas is given by the average over the inverse local depletion time distribution in star-forming regions, $\taust \equiv \Ssf/\SSFR = \langle 1/\tdep \rangle^{-1}_{\rm sf}$. Thus, if the distribution of $\tdep$ is independent of $\SH2$, so is $\taust$, regardless of the actual shape of the distribution. In our simulation, $\tdep=\tff/\epsff$, where $\epsff=\mathrm{const}$, and thus the distribution of $\tdep \propto n^{-1/2}$ is set by the density distribution of star-forming gas.
 
\begin{figure}
\centering
\includegraphics[width=\columnwidth]{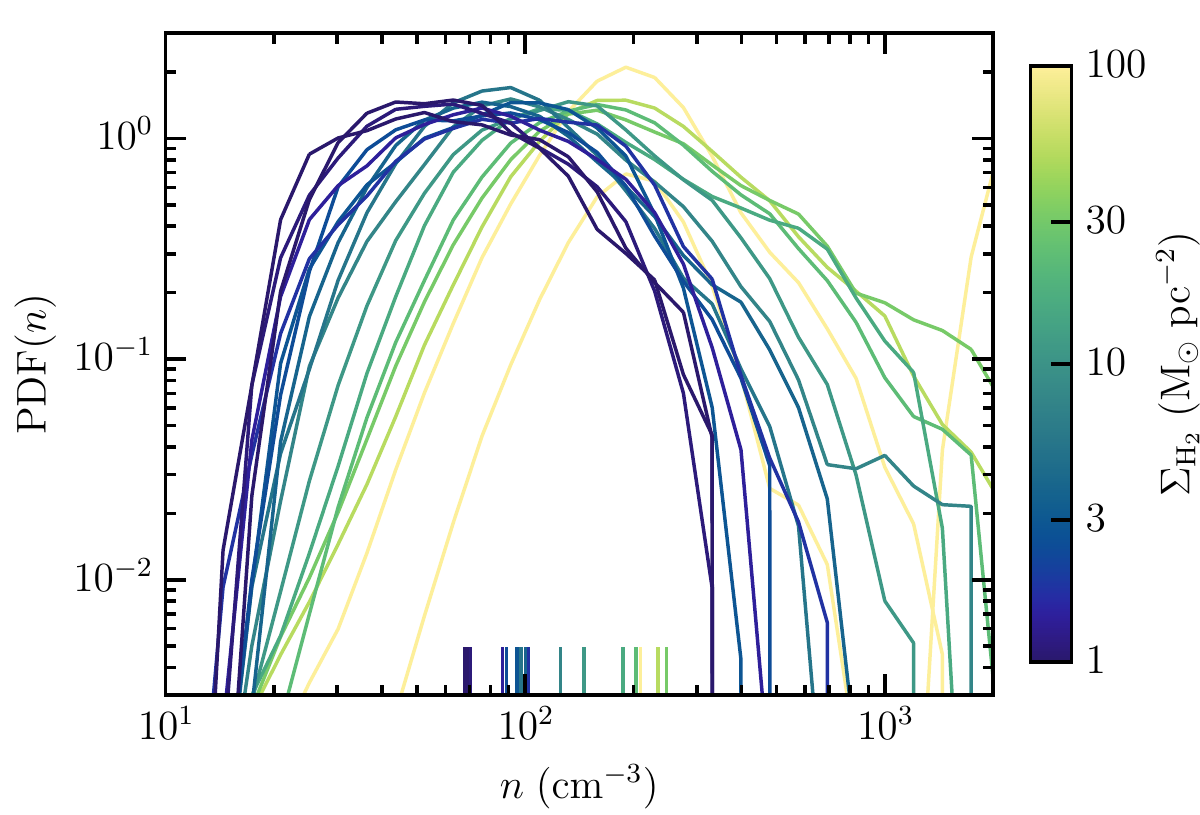}
\caption{\label{fig:pdfs} Mass-weighted number density PDFs of star-forming gas in 500~pc wide radial annuli. The line colors show the average surface density of molecular gas in a given radial annulus, $\SH2$. The vertical colored ticks indicate $\langle \sqrt{n} \rangle_{\rm sf}^2$, i.e., the average density that determines the star-forming gas depletion time in a given annulus: $\taust \propto \langle \sqrt{n} \rangle_{\rm sf}^{-1}$. To improve sampling, we accumulate the shown star-forming gas distributions between $400$ and $600 \Myr$ of disk evolution. }
\end{figure}

\begin{figure}
\centering
\includegraphics[width=\columnwidth]{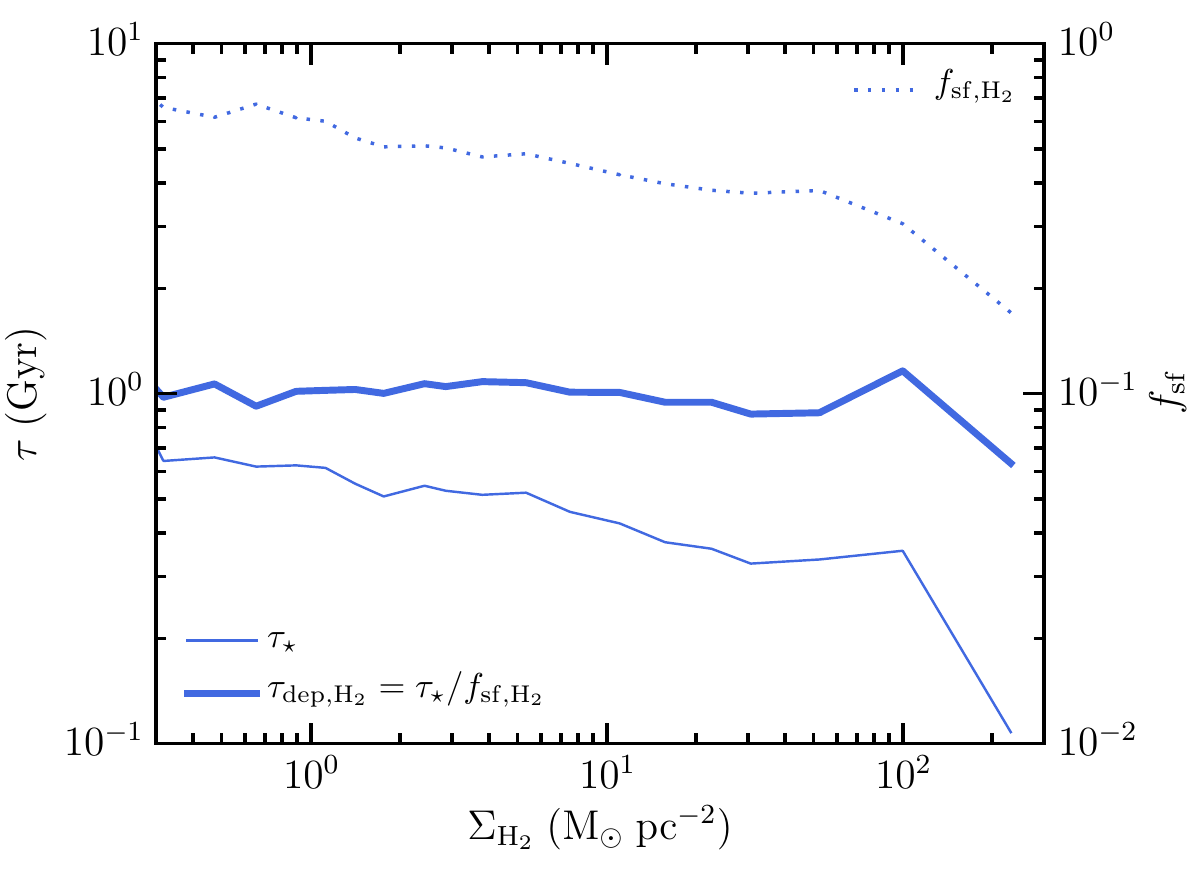}
\caption{\label{fig:slope} Dependence of molecular gas depletion time, $\tH2$ (solid line, left axis), star-forming gas depletion time, $\taust$ (thin line, left axis), and the star-forming mass fraction of molecular gas, $f_{\rm sf,H_2}$ (dotted line, right axis) on the average surface density of molecular gas, $\SH2$, in radial annuli with widths of 500 pc. The trends of $\taust$ and $f_{\rm sf,H_2}$ with $\SH2$ cancel each other out in the expression for the molecular gas depletion time, $\tglob = \taust/f_{\rm sf,H_2}$, which results in a linear KSR for molecular gas. }
\end{figure} 
 
The PDFs of the star-forming gas density in radial annuli with widths of 500 pc are shown in Figure~\ref{fig:pdfs}. The figure shows that $\SH2$ spans two orders of magnitude, while the average star-forming gas density and the shape of the PDF do vary, but the variation is quite mild. Thus, $\taust$ depends on $\SH2$, but weakly, as shown by the thin line in Figure \ref{fig:slope}. 
 
This dependence is weak for two reasons. First, the average density of star-forming gas in all of the ISM environments in our simulation is limited by the stellar feedback (at high densities) and FUV background (at  $n < 20\cc$) to a nearly constant range of $n\sim 10-10^3\ \rm cm^{-3}$, comparable to the typical densities of observed star-forming regions. Second, as in the counting argument, $\SH2$ varies mostly due to the variation of the geometric covering fraction of molecular gas, not due to the change of its density. 
 
As the dotted line in Figure~\ref{fig:slope} shows, the star-forming mass fraction, $f_{\rm sf,H_2}$, also varies weakly with $\SH2$, with a dependence on $\SH2$ similar to that of $\taust$. In Equation~(\ref{eq:KS_tdep_H2}) the weak trends of $\taust$ and  $f_{\rm sf,H_2}$ with $\SH2$ cancel, resulting in a constant $\tH2$, which explains the origin of the linear $\SSFR-\SH2$ relation.

This almost exact cancellation of the $f_{\rm sf,H_2}$ and $\taust$ trends implies that $f_{\rm sf,H_2} \propto \taust$. Some covariance of $f_{\rm sf,H_2}$ with $\taust$ is expected because $\taust$ controls the rate of feedback energy and momentum injection. As a result, longer $\taust$ results in less turbulence, i.e., smaller $\avir$ and larger $f_{\rm sf,H_2}$. This qualitative explanation, however, will need to be investigated and tested more thoroughly with simulations and will be a subject of our future study.

\vspace{1em}
\section{Discussion}
\label{sec:discussion}

The short, $\sim 10-100$ Myr, timescales of the physical processes driving the evolution of gas in the ISM (see the top panel of Figure~\ref{fig:tcdist}) indicate that the ISM is vigorously ``boiling'' when considered on the global depletion timescale. During this vigorous evolution, gas cycles between non-star-forming and star-forming stages and spends only $\tsf \sim 5 - 15 \Myr$ in the star-forming stage on each cycle (see the bottom panel of Figure~\ref{fig:tcdist}), which is consistent with the short lifetimes derived for observed GMCs \citep[e.g.,][]{Kawamura.etal.2009,Murray.2011,Schruba.etal.2017}. 

Observational estimates of the integral star-formation efficiency during a star-forming stage, defined for a given star-forming region containing a gas mass of $m_{\rm g}$ and a mass of formed young stars of $m_\star$ as $\epsint \equiv m_\star/(m_{\rm g} + m_\star)$, give $\epsint \sim 1 - 20\%$  \citep[e.g.,][]{Evans.etal.2009,Lada.etal.2010}. This fraction is even smaller in less-efficient clouds \citep[e.g.,][]{Rebolledo.etal.2015,Lee.etal.2016,Vutisalchavakul.etal.2016}. A similar range of $\epsint$ is also obtained in simulations of star cluster formation \citep{Gavagnin.etal.2017}, models of star formation in GMCs \citep[e.g.,][]{Zamora.Vazquez.Semadeni.2014}, and cosmological simulations of a Milky Way-sized galaxy that resolve the growth of globular clusters and self-consistently capture its termination by stellar feedback \citep{Li.etal.2017}. 

Such values of $\epsint$ imply that gas parcels must undergo $\Nc \sim \epsint^{-1} \sim 5-100$ cycles transitioning from the non-star-forming to star-forming state before they convert their gas into stars. This number of cycles is also consistent with the typical depletion times of star-forming gas, $\taust \equiv \langle 1/\tdep \rangle^{-1}_{\rm sf}$, and lifetimes, $\tsf$, derived for observed star-forming regions, $\Nc = \taust/\tsf$. Specifically, depletion times of gas in observed star-forming regions are estimated to be $\tdep \sim 50-500 \Myr$ \citep[e.g.,][]{Evans.etal.2009,Evans.etal.2014,Lada.etal.2010,Lada.etal.2012,Heiderman.etal.2010,Gutermuth.etal.2011,Schruba.etal.2017}; for $\tsf \sim 5 - 15 \Myr$ quoted above, these $\tdep$ give $\Nc\sim (50-500)/(5-15)\sim 3-100$. Such $\Nc$ are in the ballpark of the $\Nc$ range that we estimate for the gas in our simulations (see the bottom panel of Figure \ref{fig:tcdist}), although we note that the number of cycles in our simulation can be somewhat overestimated due to the artificially sharp threshold in the definition of star-forming gas. 

Note that specific values of $\tnsf$, $\tsf$ and $\Nc$ depend on the scale, $l$, on which the small-scale distribution of $\tdep$ is defined. Clearly, if we consider the evolution of gas parcels on the scale of protostellar cores, $\sim 0.01 \pc$, the star-forming stage of evolution will correspond to the formation of one or a handful of stars, which will consume most of the gas in a single event. The gas parcels on this scale will spend a long time in the non-star-forming stage and will consume their gas in one or a few cycles, $\Nc = 1+\xi$, where $\xi$ corresponds to the ratio of the protostellar core mass returned back to the ISM to the mass of the formed star.

The key expression of our model (Equation~\ref{eq:tglob_xi}) was derived by applying the mass conservation law to all star-forming gas in a galaxy or in a given ISM patch. Mass conservation can also be invoked to build a model for the formation, evolution, and destruction of individual GMCs \citep[e.g.,][]{Feldmann.Gnedin.2011,Zamora.etal.2012,Lee.etal.2016}.

Our model is based on mass conservation alone, and thus the overall physical explanation for long gas depletion times presented in Sections \ref{sec:model} and \ref{sec:results} does not require the assumption of dynamical equilibrium. Indeed, when a gas parcel undergoes some inherently nonequilibrium process, such as compression in a spiral arm, the parcel's depletion time will still be given by $t_{\rm dep} = (\taust/\tsf)(\tnsf + \tsf) = \taust(\tnsf/\tsf + 1)$, and therefore $t_{\rm dep}$ will be long if $\tnsf \gg \tsf$ and/or $\taust$ is long. The absence of a long-term equilibrium requirement is an essential difference of our framework from the models for the Kennicutt-Schmidt relation that rely on the assumption of self-regulation to the vertical or \citet{Toomre.1964} equilibrium state \citep[e.g.,][]{Ostriker.Shetty.2011,FaucherGiguere.etal.2013}.

In fact, our model explicitly accounts for the deviations from the equilibrium state in which $\dot{M}_{\rm sf} \approx 0$. Such deviations, along with fluctuations of other quantities that enter Equation~(\ref{eq:tglob_xi}), can be important sources of the depletion time scatter. These deviations can be substantial for individual ISM patches, which generally will not be in equilibrium, even if a galaxy as a whole is. Additional scatter can arise due to observational tracers sampling different stages of gas evolution incompletely  \citep{Kruijssen.Longmore.2014}. 

We note, however, that unlike the models of \citet{Ostriker.Shetty.2011} and \citet{FaucherGiguere.etal.2013}, our conceptual framework cannot quantitatively predict the depletion time by itself. It only elucidates how the depletion time is related to the timescales of the processes driving gas evolution. The variables through which this relation is parameterized can be either calibrated in simulations, as is done in this study, or be derived in analytical models \citep[e.g.,][]{Zamora.etal.2012}. Nevertheless, as we discuss in the following subsections, our framework is very useful for interpreting and explaining a number of puzzling facts about star formation in both observed and simulated galaxies.

\subsection{Implications for observations}
\label{sec:discussion:galaxies}

Rapid cycling of gas between non-star-forming and star-forming states explains the large discrepancy between long global depletion times of $\gtrsim 1\Gyr$ and short, $\sim 10-100\Myr$, timescales associated with the dynamical processes in the ISM. Only a small fraction of gas is converted into stars during each cycle, and therefore gas would have to go through a large number of cycles to be depleted. 

Our model also naturally explains the difference between observed local depletion times of (mostly molecular) gas in star-forming regions, $\tdep \sim 50 - 500 \Myr$, and global depletion times of both total gas, $\tglob \sim 2 - 10 \Gyr$, and molecular gas, $\tH2 \sim 1-3 \Gyr$. The global depletion times, $\tglob$ and $\tH2$, are longer than the average gas depletion time in star-forming regions, $\taust$, due to the significant fraction of time that gas spends in the non-star-forming state (see Equation~\ref{eq:tglob1}). This implies that only a fraction of total and molecular gas is forming stars at any given moment. For example, the observed values of $\tdep$ and $\tH2$ indicate that only $f_{\rm sf,H_2} \equiv \Msf/\MH2 = \taust/\tH2 \sim 5\% - 50\%$ of molecular gas is forming stars. The range of $f_{\rm sf,H_2}$ in our simulation is consistent with this estimate (see the dotted line in Figure~\ref{fig:slope}), with the non-star-forming molecular gas state corresponding to strongly turbulent cold gas. Diffuse, non-star-forming molecular gas is also observed in the Milky Way, but its mass fraction \citep[$\sim25\%$ from][]{RomanDuval.etal.2016} is a factor of $\sim 2-4$ too low to explain the discrepancy between global and local depletion times of molecular gas. Thus, a substantial fraction of non-star-forming molecular gas must be missed in such observations.

As pointed out by \citet{Kruijssen.Longmore.2014}, a model considering different evolutionary stages and corresponding chemical phases of the ISM gas can be used to interpret the dependence of the depletion time and its scatter on the averaging scale \citep[see, e.g.,][]{Schruba.etal.2017}. This dependence in observed galaxies can also be used as a stringent test of the star formation and feedback implementation in galaxy formation models.  

Our model for gas depletion time provides a natural framework for predicting and interpreting trends with galaxy properties and redshift. For instance, we show that the duration of cycles decreases with increasing surface density as $\tnsf \propto \Sg^{-0.5}$, which is accompanied by a milder but nevertheless nonnegligible decrease in $\taust$ (see Section~\ref{sec:results:KS} and the thin line in Figure~\ref{fig:slope}). This means that the observed decrease in global depletion times in high-redshift and starburst galaxies \citep[e.g.,][]{Kennicutt.1998,Bouche.etal.2007,Genzel.etal.2010,Tacconi.etal.2017} can be explained by shorter dynamical timescales, $\tnsf$, and star-forming gas depletion times, $\taust$, associated with high-density environments. In addition, the nonequilibrium state of starburst galaxies may result in short $\tglob$ due to the contribution of the $\taust/\tsfe$ term in Equation~(\ref{eq:tglob_xi}).
 
Our framework also predicts the dependence of depletion times and KSR shape on metallicity. Gas must be shielded by a certain column density in order to become cold and molecular. This column density has a corresponding number density at which such a transition occurs, as can be seen in the phase diagrams in Figure \ref{fig:phases_r}, that show the sharp change from the warm, transonic phase (yellow) to the cold, supersonic phase (blue) at $n \sim 10\cc$. At lower metallicities, both the characteristic number density and column density of the transition increase, leading to the decrease of $\fH2$, $\fsf$ and $\taust$. Thus, the overall gas depletion time, $\tglob$, increases. The mechanism of $f_{\rm sf,H_2}$ regulation by feedback-driven turbulence discussed in Section~\ref{sec:results:KS}, however, should operate regardless of the gas metallicity. Thus, we expect the KSR for molecular gas to remain linear at low metallicity. Although we also expect its normalization to be approximately independent of gas metallicity, a weak dependence is possible due to the changing properties of star-forming regions with metallicity.

The higher characteristic density of the transition at lower metallicity also results in the shift of the turnover in the KSR for total gas to higher surface densities. This shift is qualitatively similar to that predicted by the \citet{Gnedin.Kravtsov.2011} models, where star formation is tied to molecular gas. We have confirmed this explicitly by resimulating our model galaxy at a lower metallicity and will present these results in a forthcoming paper.

\subsection{Implications for galaxy formation simulations}
\label{sec:discussion:simulations}

Recent studies show that in simulations with strong feedback, the galaxy-scale star formation rate --- and hence the global depletion time --- is insensitive to the local star formation efficiency \citep[e.g.,][]{Dobbs.etal.2011,Agertz.etal.2013,Hopkins.etal.2013,Hopkins.etal.2017,Agertz.Kravtsov.2015,Benincasa.etal.2016,Orr.etal.2017}. This behavior is thought to be due to ``self-regulation'' of star formation by feedback \citep[e.g.,][]{Dobbs.etal.2011}.

Our framework naturally explains the physical mechanism of such self-regulation. Equation~(\ref{eq:tglob_xi}) accounts for feedback via the term proportional to the mass-loading factor, $\xi$. When feedback is sufficiently strong so that this term dominates, the depletion time is given by $\tglob\approx \xi\tp$, where $\tp$ is the timescale of the star-forming gas supply from the ISM. Physically, $\tp$ is not related to the local star formation efficiency, while $\xi$ is the ratio of the star-forming gas depletion time, $\taust=\langle \epsff/\tff \rangle^{-1}_{\rm sf}$, and the timescale with which feedback disperses star-forming gas, $\tmfb$ (see Section \ref{sec:model}). Timescales $\taust$ and $\tmfb$ are proportional to each other, because the rate of gas removal from star-forming regions is proportional to the rate of energy and momentum injection by feedback, which, in turn, is proportional to the star formation rate: $\Msf/\tmfb=\xi\dot{M}_\star = \xi \Msf/\taust$. Thus, in this regime, the global star formation rate, $\SFR = \Mg/\tglob \approx \Mg/(\xi\tp)$, does not depend on the local depletion time and thus on local star formation efficiency.  At the same time, $\SFR$ is inversely proportional to the strength of feedback, $\xi$, as observed in simulations with self-regulation \citep[e.g.,][]{Hopkins.etal.2017,Orr.etal.2017}. 

Note that for a given strength of feedback, the regime of self-regulation exists only for a certain range of $\epsff$ and $\taust$. For sufficiently low $\epsff$ (long $\taust)$, the second term in the sum in Equation~(\ref{eq:tglob_xi}), i.e. $\taust$, will dominate, and the global depletion time will scale with the local star formation efficiency as $\tglob \sim \taust \propto \epsff^{-1}$. Indeed, we have checked that when the feedback parameters and star formation threshold in our simulations are fixed, the global depletion time is insensitive to the variation of local efficiency for $\epsff > 1\%$, but scales as $\tglob \propto \epsff^{-1}$ when $\epsff < 1\%$.

The critical value of $\epsff$, above which $\tglob$ becomes insensitive to $\epsff$, depends on the relative contribution of the $\taust$ term in the sum of Equation~(\ref{eq:tglob_xi}), and thus depends on the factors that control the first term in the sum, in particular the feedback strength, $\xi$. Dependence of the critical $\epsff$ on feedback strength explains the results of \citet{Agertz.etal.2013} and \citet{Agertz.Kravtsov.2015}, who found that the normalization of the KSR, i.e. $\tglob^{-1}$, scales with $\epsff$ in simulations without feedback but is similar for $\epsff=1\%$ and $10\%$ when feedback is strong.

The weak sensitivity of the SFR to the value of $\epsff$ in the regime of strong feedback does not mean that this value is not important. For example, the mass fraction of star-forming gas in this regime scales inversely with $\epsff$: $\fsf \equiv \Msf/\Mg = \taust/\tglob \propto \epsff^{-1}$. Thus, an incorrect $\epsff$ will result in an incorrect mass fraction and an incorrect spatial distribution of star-forming gas. A similar conclusion was drawn by \citet{Hopkins.etal.2013b}, who suggested that the amount of dense gas probed by HCN decreases with increasing star formation efficiency and feedback strength.

The mass fraction of star-forming gas and its distribution is also a nontrivial function of the feedback strength and the specific mix of processes that define it \citep[e.g.,][]{Hopkins.etal.2013b,Butler.etal.2017}. Stronger feedback shortens the time that gas spends in the star-forming state, $\tsf$, and, therefore, decreases the mass fraction of star-forming gas: $\fsf \equiv \Msf/\Mg \sim \tsf/(\tnsf+\tsf)$. Besides, strong feedback shapes the overall gas PDF and may increase timescales $\tnsf$, which also results in a decrease of $\fsf$. This happens, for example, when a gaseous disk is stabilized by a pressure corresponding to a slowly dissipating energy directly injected by feedback \citep[e.g.,][]{Agertz.etal.2013,Benincasa.etal.2016}. 

Finally, $\tnsf$, $\tsf$ and $\fsf$ depend on the overall adopted definition of star-forming gas. Usually, the boundary between star-forming and non-star-forming gas is established by a set of thresholds in various gas properties, such as gas density, molecular fraction, virial parameter $\avir$, etc. Our results indicate that these thresholds should be carefully chosen, because if thresholds grossly misidentify the star-forming gas, the values of $\epsff$ and feedback strength may need to compensate for the wrong choice to get reasonable $\tsf$ and $\fsf$ values. It is not yet clear whether such compensation is possible in general, but, in any case, wrong thresholds may drive the $\epsff$ and feedback parameters to incorrect values. 

In addition, the choice of the thresholds defining star-forming gas indirectly affects the efficiency of feedback. When more gas is designated as star-forming for a given fixed star formation rate, the feedback mass-loading factor, $\xi$,  will have to be larger, i.e., feedback has to be stronger. The specific choice of the thresholds matters as well. For example, when star-forming gas is defined using a density threshold, $n_{\rm sf}$, feedback has to drive star-forming gas to lower densities, $n<n_{\rm sf}$, in order to shut down star formation. If star-forming gas is defined using a threshold in the virial parameter instead, $\avir$ can be increased to values larger than the threshold very quickly by injecting thermal or turbulent energy without changing the gas density significantly.

Thus, the framework presented in this paper implies that to get the value of $\fsf$ and depletion time, $\tglob = \taust/\fsf$, and the distribution of star-forming gas correctly, the overall definition of the star-forming gas, its star formation efficiency, and the strength of stellar feedback are all important and should all be modeled carefully. Ideally, the modeling choices should be based on solid physical ground or subgrid models calibrated on higher-resolution simulations. For example, recent studies show that local $\epsff$ can be modeled using the results of high-resolution simulations of turbulent star-forming gas \citep{Padoan.etal.2012,Padoan.etal.2017,Federrath.2015} and a subgrid model for turbulence calibrated on turbulence simulations \citep[e.g.,][]{Braun.etal.2014,Braun.Schmidt.2015,Semenov.etal.2016,Li.etal.2017}.

Interestingly, the results presented in Section~\ref{sec:results:KS} indicate that feedback can break the direct connection between the small- and large-scale relations between gas density and SFR. Specifically, our simulations assume $\rhoSFR \propto \rho^{1.5}$ on the scales of individual cells. However, as can be seen in Figures~\ref{fig:KS} and \ref{fig:slope}, on kiloparsec scales, the relation between the surface density of molecular gas and star formation rate is almost linear. This result provides a counterexample to the arguments that in simulations, the slope of the Kennicutt-Schmidt relation on kiloparsec scales simply reflects the assumed density dependence of the local SFR on small scales \citep{Schaye.DallaVecchia.2008,Gnedin.etal.2014}. Our analysis shows that the linear relation arises because only a fraction of cold, dense gas is forming stars and its density PDF does not scale self-similarly with the large-scale surface density of molecular gas, $\SH2$. Variation in covering fraction allows $\SH2$ to vary significantly, while the average density of star-forming gas and the corresponding depletion time vary only weakly. In addition, self-regulation by feedback ties the fraction of molecular gas that is forming stars and the depletion time of star-forming gas, so that their trends with $\SH2$ are similar. These trends cancel out resulting in an almost constant $\tau_{\rm dep, H_2}$ and linear $\dot{\Sigma}_\star-\SH2$ relation. 

The emergence of the large-scale KSR via nontrivial, nonlinear effects of the star formation-feedback loop motivates efforts to model such processes in high-resolution simulations rather than tuning the star formation prescription to produce a particular large-scale relation. 

Our model also sheds some light on the importance of the so-called ``early feedback'' --- a collective name for the energy and momentum injection by young stars during the first 3 Myr of the life of a stellar population, before first supernovae explode. The time lag between the formation of stellar particles and the onset of feedback is important when $\taust$ is comparable to the lag. In this case, a substantial fraction of the gas mass can be converted into stars in the first 3 Myr before supernova feedback can limit star formation. When $\taust$ is long, the global depletion time becomes less sensitive to the presence or absence of stellar feedback during the first 3 Myr of the star-forming stage.

\section{Summary and conclusions}
\label{sec:conclusions}

We present a simple and intuitive physical model that elucidates why gas depletion times in galaxies are long compared to the timescales of the processes driving the evolution of the interstellar medium. We show that the depletion time is long not because some bottleneck in the formation of star-forming regions imposes a long evolutionary timescale, but because only a small fraction of the gas mass is converted into stars during a single star-forming stage in the evolution of a gas parcel. This fraction is small due to both the short duration of the star-forming stage, as dynamical processes and stellar feedback efficiently disperse star-forming regions, and the low intrinsic star formation efficiency of dense molecular gas. A gas parcel thus must go through many cycles transitioning between non-star-forming and star-forming states before it becomes converted into stars. Hence, even though the duration of each cycle can be short, the global depletion time is long because the number of cycles is large.  

Furthermore, the difference between the global and local depletion times of molecular gas in our model arises because not all of the molecular gas is actively forming stars. Non-star-forming molecular gas appears naturally if local star formation efficiency is a strong function of the virial parameter of a region, while the molecular fraction of gas is set by its ability to shield against FUV radiation and is a function of mainly gas number density and metallicity. 

We illustrate our model using the results of an isolated $L_*$-sized disk galaxy simulation that reproduces the observed Kennicutt-Schmidt relation for both molecular and atomic gas. We discuss the predictions of our model for the dependence of the global depletion time on properties of observed galaxies and on the parameters of star formation and feedback recipes in galaxy simulations. In particular, our model explains the weak sensitivity of the global star formation rate to the assumed local star formation efficiency reported in several recent numerical studies and the physics of this ``self-regulation'' (see Sections \ref{sec:model} and \ref{sec:discussion:simulations}). 

Additional results and conclusions can be summarized as follows. 
\begin{itemize}
\item[1.] Analysis of our simulation shows that the properties of gas parcels in the ISM evolve on timescales of $\sim 10-100\Myr$ under the influence of compression by the spiral arms, ISM turbulence, and SNe-driven shocks. The relative importance of these processes varies with galactocentric radius, $R$, and average surface density. At $R \gtrsim 5$ kpc, the evolution from a warm, diffuse state to a dense, cold phase is driven mainly by compression in the spiral arms, while SNe-driven shocks and large-scale ISM turbulence dominate at smaller radii. 

\item[2.] During an evolutionary cycle, gas spends most of the time in the non-star-forming state, $\tnsf > \tsf$, whereas the time spent in the star-forming state, $\tsf$, is limited by stellar feedback and dynamical processes to $\tsf \sim 5 - 15 \Myr$.  We find that the median $\tnsf$ varies with gas surface density as $\tnsf \propto \Sg^{-0.5}$.

\item[3.] On the resolution scale of our simulation, $40$ pc, the typical range of densities in star-forming regions is limited to $n \sim 10 - 10^3 \cc$ by the interstellar FUV background and stellar feedback. We find that the resulting depletion times of star-forming gas $\taust \sim 300 - 500 \Myr$ are consistent with the depletion times estimated for observed GMCs on these scales and exhibit only weak trends with the surface densities of total and molecular gas.  

\item[4.] The distributions of depletion times and lifetimes of star-forming regions in our simulations imply that a typical gas parcel has to undergo $5-100$ cycles transitioning between non-star-forming and star-forming states before converting its mass into stars. 

\item[5.] On kiloparsec scales, our simulation produces a nearly linear relation between the surface density of H$_2$ and surface density of star formation rate, i.e., the molecular Kennicutt-Schmidt relation, even though a nonlinear local relation, $\rhoSFR \propto \rho^{1.5}$, is adopted for star-forming gas in simulation cells. We show that the linear relation emerges due to stellar feedback, which shapes the gas density PDF making it non-self-similar and establishes a correlation between the local depletion time of the star-forming gas and the fraction of molecular gas that is in the star-forming state.  
\end{itemize}

The model for the global depletion timescale presented in this paper is a generic framework that can be applied not only to galaxies as a whole but also to individual ISM patches with sizes ranging from $\sim$ kiloparsec to a typical size of star-forming regions, $\sim 10$ pc. It can also be used to predict and interpret trends of gas depletion time with the ISM properties and redshift. 

As a final comment, we note that in the context of galaxy evolution over cosmological timescales, the actual gas depletion time is often considered to be unimportant. For example, galaxies from the star-forming sequence are predicted to form stars at the rates regulated by gas accretion and gas loss in winds, because gas depletion times in such galaxies are short compared to other relevant timescales \citep[e.g.,][]{Bouche.etal.2010,Dave.etal.2012,Lilly.etal.2013}. Note, however, that dwarf galaxies and galaxies from the green valley consume gas on extremely long timescales of $\gtrsim 5 - 10 \Gyr$, and, therefore, their depletion times do affect their evolution. Moreover, at $z \gtrsim 5-6$, when the age of the universe is $\lesssim 1$ Gyr, the Gyr-long gas depletion times become comparable to the cosmological evolution timescale and will therefore play an important role in controlling the SFR during the early stages of galaxies evolution \citep[e.g.,][]{Dekel.Mandelker.2014,Peng.Maiolino.2014}. The framework for modeling gas depletion time presented in this paper thus opens a way to refine theoretical models of galaxy formation in this regime, which is particularly important in the upcoming era of the JWST. 

\acknowledgments

We would like to thank the referee, Mark Krumholz, for a constructive report and many detailed comments that have improved the paper. We thank Cameron Liang, Claude-Andr\'e Faucher-Gigu\`ere, Robert Feldmann, Romain Teyssier,  Simon Lilly, and Marcella Carollo for useful discussions. We are also grateful to Oscar Agertz, Camille Avestruz, Benedikt Diemer, Oleg Gnedin, Hui Li, Philip Mansfield, Enrique V{\'a}zquez-Semadeni, and Tony Wong, whose valuable comments and suggestions helped to improve our paper. We also thank Adam Leroy for sharing the data from \citet{Leroy.etal.2013} that we used in Figure~\ref{fig:KS}. This work was supported by a NASA ATP grant NNH12ZDA001N, NSF grant AST-1412107, and by the Kavli Institute for Cosmological Physics at the University of Chicago through grant PHY-1125897 and an endowment from the Kavli Foundation and its founder, Fred Kavli. The simulation and analyses presented in this paper have been carried out using the Midway cluster at the University of Chicago Research Computing Center, which we acknowledge for support.

\bibliographystyle{aasjournal}
\bibliography{}

\end{document}

%% file: citation_fix.tex
\usepackage{etoolbox}
\makeatletter
\patchcmd{\NAT@citex}
  {\@citea\NAT@hyper@{\NAT@nmfmt{\NAT@nm}\NAT@date}}
  {\@citea\NAT@nmfmt{\NAT@nm}\NAT@hyper@{\NAT@date}}
  {}
  {}
\patchcmd{\NAT@citex}
  {\@citea\NAT@hyper@{%
     \NAT@nmfmt{\NAT@nm}%
     \hyper@natlinkbreak{\NAT@aysep\NAT@spacechar}{\@citeb\@extra@b@citeb}%
     \NAT@date}}
  {\@citea\NAT@nmfmt{\NAT@nm}%
   \NAT@aysep\NAT@spacechar%
   \NAT@hyper@{\NAT@date}}
  {}
  {}
\patchcmd{\NAT@citex}
  {\@citea\NAT@hyper@{%
     \NAT@nmfmt{\NAT@nm}%
     \hyper@natlinkbreak{\NAT@spacechar\NAT@@open\if*#1*\else#1\NAT@spacechar\fi}%
       {\@citeb\@extra@b@citeb}%
     \NAT@date}}
  {\@citea\NAT@nmfmt{\NAT@nm}%
   \NAT@spacechar\NAT@@open\if*#1*\else#1\NAT@spacechar\fi%
   \NAT@hyper@{\NAT@date}}
  {}
  {}
\makeatother